\documentclass[twocolumn,trackchanges,times]{aastex631}

\usepackage{graphicx,amsmath,amsfonts,amssymb,slashed,hyperref}
\usepackage{lipsum}
\let\tablenum\relax
\usepackage{siunitx}
\usepackage{xcolor}
\usepackage{textcomp}
\usepackage{url}
\usepackage{natbib}

\usepackage{color}

\usepackage{epstopdf}
\usepackage{gensymb}
\usepackage{hyperref}
\usepackage{longtable}
\usepackage{subfigure}
\usepackage[T1]{fontenc}

\newcommand*{\teff}{$T_{\rm eff}$}
\newcommand*{\logg}{$\log~g$}
\newcommand*{\afe}{[$\alpha$/Fe]}

\newcommand*{\feh}{[Fe/H]}
\newcommand*{\kms}{km s$^{-1}$}
\newcommand*{\zmax}{$Z_{\rm max}$}

\newcommand*{\vphi}{$V_{\rm \phi}$}

\newcommand*{\z}{$|Z|$}

\newcommand*{\tc}{$T_{\rm C}$}
\newcommand*{\vz}{$V_{\rm Z}$}
\newcommand*{\re}{$R_{\oplus}$}
\newcommand*{\rmean}{$R_{\rm M}$}
\newcommand*{\np}{$N_{\rm P}$}

\DeclareUnicodeCharacter{2212}{-}

\begin{document}

\title{Connections between Planetary Populations and the Chemical Characteristics of their Host Stars}

\shorttitle{Connections between Planetary Populations and the Chemistry of Host Stars}
\shortauthors{Yun et al.}

\author[0009-0004-1035-3309]{Sol Yun}
\affiliation{Department of Astronomy, Space Science, and Geology, Chungnam National University, Daejeon 34134, South Korea; yunsol719@gmail.com}

\author[0000-0001-5297-4518]{Young Sun Lee}
\affiliation{Department of Astronomy and Space Science, Chungnam National University, Daejeon 34134, South Korea; youngsun@cnu.ac.kr}
\affiliation{Department of Physics and Astronomy, University of Notre Dame, Notre Dame, IN 46556, USA}

\author[0000-0002-6411-5857]{Young Kwang Kim}
\affiliation{Department of Astronomy and Space Science, Chungnam National University, Daejeon 34134, South Korea}

\author[0000-0003-4573-6233]{Timothy C. Beers}
\affiliation{Department of Physics and Astronomy, University of Notre Dame, Notre Dame, IN 46556, USA}
\affiliation{Joint Institute for Nuclear Astrophysics -- Center for the Evolution of the Elements (JINA-CEE), USA}

\author{Berfin Togay}
\affiliation{Department of Astronomy, Space Science, and Geology, Chungnam National University, Daejeon 34134, South Korea}

\author[0000-0001-7277-7175]{Dongwook Lim}
\affiliation{Department of Astronomy, Yonsei University, Yonsei-ro, Seodaemun-gu Seoul, 03722, South Korea}

\begin{abstract}

Chemical anomalies in planet-hosting stars (PHSs) are studied in order to assess how the planetary nature and multiplicity affect the atmospheric chemical abundances of their host stars. We employ APOGEE DR17 to select thin-disk stars of the Milky Way, and cross-match them with the Kepler Input Catalog to identify confirmed PHSs, which results in 227 PHSs with available chemical-abundance ratios for six refractory elements. We also examine an ensemble of stars without planet signals, which are equivalent to the selected PHSs in terms of evolutionary stage and stellar parameters, to correct for Galactic chemical-evolution effects, and derive the abundance gradient of refractory elements over the condensation temperature for the PHSs. Using the Galactic chemical-evolution corrected abundances, we found that PHSs do not show a significant difference in abundance slope from the stars without planets. Furthermore, we examine the depletion trends of refractory elements of PHSs depending on total number of planets and their types, and find that the PHSs with giant planets are more depleted in refractory elements than those with rocky planets. Among the PHSs with rocky planets, the refractory-depletion trends are potentially correlated with the terrestrial planets’ radii and multiplicity. In the cases of PHSs with giant planets, sub-Jovian PHSs demonstrated more depleted refractory trends than stars hosting Jovian-mass planets, raising questions on different planetary-formation processes for Neptune-like and Jupiter-like planets.

\end{abstract}


\section{Introduction} \label{sec:intro}

A star and its planets are thought to have formed from the same molecular cloud. The composition of a star could influence the protoplanetary disk where planets form, and the accreting material from the protoplanetary disk onto the host star may be chemically imprinted in the atmosphere of the star. These interactions may result in an intricate relationship between the chemical composition of a star and the formation and evolution of its planets. Various approaches have been conducted to find such connections between the chemical composition of a planet-hosting star (PHS) and its planet formation and architecture. Notably, the amount of heavy elements in a stellar atmosphere, often characterized by metallicity ([Fe/H]), has received attention due to its expected implications for the occurrence and properties of planetary companions \citep[e.g.,][]{Gonzal1997, Heiter, santos, Fischer, Udry_metal_host, Adibe_2012}.

\cite{Mel} first reported, based on an analysis of 11 Solar twin stars, that the Sun's refractory elements (Mg, Al, Si, etc.), which have condensation temperatures (\tc) over 1200\,K, are relatively depleted compared to those of the volatile elements (C, N, O, etc.). They suggested that refractory depletion is correlated with the presence of terrestrial planets; the material to form the rocky planet readily incorporates elements with high \tc\ in the solid phase, while the low \tc\ volatile elements likely remain in the gas phase. This was further supported by \cite{Cham_TEABUND}, who claimed that the deficit of the refractory elements in the solar photosphere could account for roughly four Earth masses of terrestrial material. These studies have triggered numerous efforts to identify differences in the chemical abundances between PHSs and non-PHSs (NPHSs) using chemical-abundance trends as a function of condensation temperature \citep[e.g.,][]{Rami_2009, Gon2011, Gon2013_nosignal, Adi2014, Niss2015, Bedell, Liu2020, Niba, Tau}.

When investigating the chemical-abundance trends of a star, one also needs to carefully take into account how Galactic chemical evolution (GCE) affects the chemical-abundance pattern. \cite{Bedell} identified the relative depletion of refractory elements from an analysis of 26 elements for 79 Solar twins. At the same time, they recognized that some of the depletion trend may be caused by GCE. They further demonstrated that the depletion trend is real, after carefully correcting for the GCE effects in their Sun-like stars. Such study indicates that the GCE correction is a vital step when comparing stars with different ages but similar metallicity and temperatures \citep[e.g.,][]{GCE, Spina2016_elemage3, Spina2018, Pignatari_GCE}. Other studies \citep{Adi2014, Niss2015, Spina2016_elemage3} also show that a careful selection of the comparison stars with similar stellar parameters is crucial, due to the presence of correlations between stellar parameters and chemical-abundance patterns before application of the GCE correction. As the GCE correction is non trivial, in order to minimize systematic uncertainties resulting from GCE (and adopted stellar models) several studies focused their attention on wide binaries whose evolutionary states are nearly identical, so that these effects are not significant for the analysis \citep[e.g.,][]{WB_Liu_2014, WB_Tucci, WB_Rami, WB_Saffe, WB_teske, WB_Oh, WB_Liu2021, Biazzo}.

In spite of great progress, the chemical anomalies observed in the Sun relative to Solar twins has not been consistently confirmed as a unique feature in other studies \citep[e.g.,][]{Gonz_2010, Mel_2012_solar_simil, Gon2013_nosignal, Niss2015, Liu2020, elemage4_Nissen2020, Niba}. Additionally, studies that examined the relationship between stellar-chemical anomalies and planet formation were not successful in detecting a definitive signal of such a relationship, and triggered more controversy \citep{Adi2014, Liu2020, Tau, Behmard_2023_APG17}. \cite{Niba}, for instance, found two groups of stars: chemically depleted and non-depleted ones, by introducing a likelihood-based approach to determine elemental abundances in Solar analog stars identified in Apache Point Observatory Galactic Evolution Experiment Data Release 16 \citep[APOGEE DR16;][]{APG16}. They agreed on the refractory depletion of the Sun with \cite{Bedell}, arguing that the Sun is one of the stars in their refractory-depleted group. 

Similarly, \cite{Liu2020} analyzed 16 PHSs and 68 comparative NPHSs, following the methods of \cite{Bedell}, and found various abundance-\tc\ trends among PHSs without a clear signature of the depletion of the refractory elements for the PHSs. They  suggested that a range of possible planet-induced effects are responsible. \cite{Behmard_2023_APG17} carried out a study on the chemical dissimilarity of 12 elements between 130 known/candidate Kepler Objects of Interest (KOIs) of PHSs and so-called doppelg\"angers stars, selected based on the proximity of four parameters: effective temperature (\teff), surface gravity (\logg), \feh, and [Mg/H] in high-resolution near-infrared spectra from APOGEE DR17 (Abdurro'uf et al.\ \citeyear{APG17}). They reported that the median intrinsic dispersion between the KOI and doppelg\"anger samples were consistently under 0.05 dex, and argued that there are no noticeable signatures for different chemical-abundance patterns associated with PHSs.

Based on the studies to date, a consensus on the connection between the chemical properties of a PHS and its planet-formation history has yet to be achieved, nor have any clear differences in the chemical properties between PHSs and NPHSs been identified. This leads us to investigate how the planet population (number of planets and their types; often referred to as the ``architecture" of a planetary system) affects the chemical-abundance patterns of its host star, which has not been thoroughly explored previously. To achieve this goal, we employ a large sample of stars from APOGEE DR17 with well-determined stellar parameters and chemical abundances. 

This paper is arranged as follows.  In Section \ref{sec:2}, we identify dwarfs and sub-giants that are likely members of the Galactic thin disk, based on  their chemical and kinematic properties. We confirm the stars that are PHSs and NPHSs by cross-matching the APOGEE sample with the Kepler Input Catalog (KIC) and Kepler catalogs, respectively. We also identify likely twin stars within these sub-samples. We apply GCE corrections to both sub-samples in Section \ref{sec:3}. Section \ref{sec:4} presents the abundance trends as a function of \tc\ for the PHSs. In Section \ref{sec:5}, we present the results of our detailed analysis on the overall trends of the abundance slopes between the terrestrial planets and the Jovian planets. In addition, we consider the dependency of chemical-abundance depletion on the nature of the host's planet population. We discuss the implications of our findings regarding planet architecture, along with those of other studies, in Section \ref{sec:6}. A summary and conclusions are provided in Section \ref{sec:7}.

\section{Data} \label{sec:2}
\subsection{Identification of Thin-disk Stars} \label{subsec:2.1}
To identify a star with planets, we first selected likely thin-disk stars of the Milky Way (MW) of various chemical abundances, using the large spectroscopic survey data from APOGEE DR17 (Abdurro'uf et al.\ \citeyear{APG17}). APOGEE DR17 provides high-resolution ($R \sim$ 22,500) spectra in the $H$-band across the MW. The stellar parameters and abundance ratios for numerous chemical elements in these spectra are delivered by the APOGEE Stellar Parameters and Abundances Pipeline (ASPCAP; Garc{\'\i}a P{\'e}rez et al.\ \citeyear{ASPCAP}).

We computed the space-velocity components and orbital parameters of the APOGEE stars using astrometric data from Gaia Data Release 3 (Gaia DR3; Gaia Collaboration et al.\ \citeyear{GAIADR3}) and a St\"ackel-type potential, as employed in several previous studies \citep[e.g.,][]{Chiba_stackel, Kim_stackel, Kim2021, Kim2023, Lee2019, Lee2023,  Kang2023}. For these calculations, we adopted a local standard of rest velocity ($V_{\mathrm{LSR}} = 236$ km s$^{-1}$; Kawata et al.\ \citeyear{LSR}), a Solar location of $R_{\odot} = 8.2$ kpc and $Z_{\odot} = 20.8$ pc from \cite{solar_position_r} and \cite{solar_position_z}, respectively, and a Solar peculiar motion ($U, V, W)_{\odot} = (-11.10, 12.24, 7.25)$ km s$^{-1}$ \citep{solar_motion}. Among the orbital parameters, we derived the maximum distance ($Z_{\mathrm{max}}$) from the Galactic plane, which is often used as one of the conditions to select disk stars. We applied the --0.017 mas systematic offset reported in Gaia DR3 \citep{Calibrationgdr3} when calculating the distances.

To select the disk-star population, we imposed the following criteria: 3100 $\leq$ \teff\ $\leq$ 6500\,K, \logg\ $\geq$ 3.5, [Fe/H] $>$ --1.2, $7 \leq R \leq 11$ kpc, $d$ $<$ 4 kpc, \zmax\ $<$ 3 kpc, signal-to-noise ratio (S/N) $>$ 50, \vphi\ $>$ 50 \kms, \vz\ $<$ 100 \kms, and relative parallax error less than 20\%. $R$ is the distance from the Galactic center projected onto the Galactic plane, $d$ is the heliocentric distance, \vphi\ is the rotational velocity, and \vz\ is the vertical velocity component. Then, the identification of likely thin-disk members was conducted following the methodology of \cite{HanD}, who used both chemical abundances ([Fe/H] and \afe) and kinematics to derive the membership probability of each star. Note that, in this study, we attempt to analyze PHSs in the Galactic thin disk, because the thin-disk stars generally do not exhibit chemical peculiarities, unlike the thick-disk and halo stars. Additionally, we excluded in our program stars the likely accreted stars, whose origins and natures are not similar to the canonical disk stars. Following the chemical criteria of \cite{accretion}, we identified 30,265 stars with [Fe/H] $<$ --0.4 and [Al/Fe] $<$ --0.075 as accreted objects. These various selection criteria left us with 101,393 dwarf stars likely to belong to the MW's thin disk.

\subsection{Chemical Elements} \label{subsec:2.2}
We now characterize PHSs through examination of their chemical-abundance trends. Among the abundance ratios of 18 elements made available in APOGEE DR17, we employed those with elemental abundances having uncertainties that are sufficiently small to achieve the sensitivity required to detect abundance trends as a function of \tc\ \citep[]{lodd} in a star; less than 0.03 dex, as demonstrated in previous studies \citep{Adi2012, Bedell_precision, Liu2020, Niba}. We chose six refractory elements (Mg, Al, Si, Ca, Mn, and Ni) with the following conditions in mind. First, their \tc\ is over 1000 K, the lowest temperature of the refractory elements \citep{Mel, Rami_2009, Liu2020}. Secondly, their absolute abundance errors are less than 0.02 dex, with no flags raised in $X$\textunderscore $FE$\textunderscore $FLAG$ for each element. Thirdly, they are key elements for studying and testing planet-formation scenarios \citep{terrestrial, Wang}. These conditions allowed us to select 47,988 stars from our thin-disk candidate stars with well-determined abundances of Mg, Al, Si, Ca, Mn, and Ni. We did not include the abundances for the volatile elements, because their relatively larger abundance errors in APOGEE DR17 makes their abundance-\tc\ trends uncertain.

\subsection{Planet-Hosting Stars and their Twins} \label{subsec:2.3}
\subsubsection{Selection of Planet-Hosting Stars} \label{subsec:2.3.1}

We identify PHSs by cross-matching our selected thin-disk stars from APOGEE DR17 with the KIC Data Release 10 (Brown et al. \citeyear{KIC}) and the cumulative catalog of KOI in NASA Exoplanet Archive\footnote[8]{\url{https://exoplanetarchive.ipac.caltech.edu/}}. Within the catalog provided on the NASA Exoplanet Archive website, we used the PHS data with ``\textbf{koi\textunderscore disposition} = Confirmed'' to avoid any confusion between the other stellar and planetary signals from various sources. Furthermore, we compiled the planetary parameters, such as the planetary radius, of the selected PHS samples from the confirmed planet catalog (published on October 18, 2023) with ``\textbf{default\textunderscore flag} = 1'', resulting in a sample of 283 stars hosting 423 planets.

Similarly, we selected 2,356 comparison stars by cross-matching with the Kepler data without planetary signatures, which we take to be NPHSs, after excluding 142 stars with candidate flags. Once again, we cross-matched our data with the TESS Object of Interest (Guerrero et al. \citeyear{TOI}) candidate list to make sure of the absence of any possible planetary signals in our comparison sample. This sample is used for selecting the twin stars of the PHSs, as described below.

\subsubsection{Selection of Twins for Planet-Hosting Stars} \label{subsec:2.3.2}

To search for chemical distinctions among PHSs, we require a sample of stars without planets to compare with, and used to correct for GCE effects. It is also necessary for them to be in similar evolutionary stages and metallicities to the PHSs. However, it is extremely difficult and time-consuming to identify the NPHSs that are exact counterparts of the PHSs. Following the work of \cite{Niba} and  \cite{Behmard_2023_APG17}, we employ $twins$ of the PHSs (TPHSs), which are similar spectral types and luminosity classes to their counterpart PHSs. 

The definition of a stellar twin is diverse in the literature. In order to demonstrate the depletion of the Sun's refractory elements with respect to volatile elements, \cite{Mel} considered as Solar twins the stars with \teff, \logg, and [Fe/H] are less than 75\,K, 0.1 dex, and 0.07 dex, respectively from those of the Sun, \teff\ = 5777 K, \logg\ = 4.44, [Fe/H] = 0.0. Similarly, \cite{Rami_2009} and \cite{Bedell} adopted the criteria of $\Delta$100\,K in \teff, $\Delta$0.1 dex in \logg, and $\Delta$0.1 dex in \feh\ to select Solar twins, while \cite{Berke} used the condition of $\Delta$100\,K in \teff, $\Delta$0.2 dex in \logg, and $\Delta$0.1 dex in [Fe/H]. In contrast, \cite{Liu2020} considered only effective temperature and metallicity to increase the number of Solar analogs, while \cite{Behmard_2023_APG17} introduced [Mg/H] as an additional criterion to evaluate the similarities among their stars.

Having reviewed the selection criteria used in previous studies, among the NPHSs identified in Section \ref{subsec:2.3.1}, we sorted out the TPHSs by adopting the "twin regions" of $\Delta$100\,K in \teff, $\Delta$0.1 dex in \logg, and $\Delta$0.1 dex in [Fe/H] from the parameters of each PHS to closely match their evolutionary states. Each PHS in our sample has, on average, 43 TPHSs within their twin regions. We further confirmed that the selected TPHSs have close proximity to the target PHS by enforcing the total distance of the parameters to the minimum in a $\chi^{2}$ fashion, as employed by \cite{Behmard_2023_APG17}:

\begin{equation} \label{Chi-square}
\begin{split}
D^2 &= \left(\frac{T_{\text {eff}P}- T_{\text {eff}T}}{\sqrt{\sigma_{T_{\text {eff}P}}^2+\sigma_{T_{\text {eff}T}}^2}}\right)^2+\left(\frac{ \log g_{P} - \log g_{T}}{\sqrt{\sigma_{\log gP}^2+\sigma_{\log gT}^2}}\right)^2 \\
&+ \left(\frac{[\mathrm{Fe} / \mathrm{H}]_{P}-[\mathrm{Fe} / \mathrm{H}]_{T}}{\sqrt{\sigma_{[\mathrm{Fe} / \mathrm{H}] P}^2+\sigma_{[\mathrm{Fe} / \mathrm{H}] T}^2}}\right)^2,
\end{split}
\end{equation}

\noindent where $P$ indicates a PHS, and $T$ is the twin of a PHS. We only used the first 20 TPHSs with the smallest $D^{2}$ value for further analysis.

Of 283 PHSs, 40 stars have less than 20 TPHSs, with 12 TPHSs on average within their twin regions. When relaxing the selection criteria of TPHS to include up to 20 TPHSs for those PHSs, their stellar parameters significantly vary, ranging from $\Delta$0.6 dex in [Fe/H] and $\Delta$1.0 dex in \logg, which is far from our definition of the twins. We thus excluded these 40 stars from our analysis, leaving 243 PHSs hosting 397 planets in total.

\section{Corrections for Galactic Chemical Evolution} \label{sec:3}

Previous studies recognized that corrections for GCE are a crucial step for accurately evaluating the abundance trend of PHSs in the abundance-\tc\ plane \citep{Adi2014, Spina2016_elemage3, Bedell}. GCE refers to how chemical elements formed and were distributed in the MW \citep{GCE, Spina2016_elemage3, Pignatari_GCE}. The formation rate and the amount of elements created vary depending on the star-formation rate and initial mass function. For instance, our target elements, Mg, Si, and Ca, which are often called ``alpha elements'', are mostly created during massive-star evolution and distributed into the interstellar medium (ISM) by Type II supernovae (SNe), while Mn and Ni (iron-peak elements) are formed during the explosion of Type Ia SNe, an event involving relatively low-mass stars. Since the timescales for these progenitors differ, depending on how the ISM has been chemically enriched, a star formed out of the ISM exhibits a distinctive chemical-abundance pattern that can vary from star-to-star according to its age. To elucidate the chemical trend that represents the stellar-birth environment, we need to remove the stochastic variation of chemical abundances caused by chemical evolution. This is known as the ``GCE correction.''

Stellar age is a crucial parameter for the GCE correction. In this regard, some studies \citep{Bedell, Liu2020} excluded alpha-enhanced, old thick-disk stars from estimating GCE effects. However, we do not exclude stars older than 10 Gyr, because our samples are selected as likely thin-disk stars based on their chemical and kinematic properties. Moreover, we did not find old stars whose abundances of the six refractory elements significantly deviated from the main abundance trend of relatively younger stars, which was the primary reason for excluding the old stars in other studies. Nonetheless, to ensure the accuracy of our stellar ages, we eliminated the stars whose age errors are larger than 3 Gyr in the catalog provided by \cite{ageann}. The stellar ages reported in this catalog are determined by the astroNN \footnote[9]{\url{https://github.com/henrysky/astroNN}} package, and it includes asteroseismic ages from APOKASC-2 \citep{APOKA2} and low-metallicity asteroseismic ages \citep{APOKA_age}. This additional age constraint leaves us with 227 PHSs and 2344 TPHSs with well-determined ages. We note that TPHSs are repetitively used for each PHS.

Most of previous studies corrected for GCE by subtracting the abundance trend of each element with respect to the stellar age \citep{Niss2015, elemage2_nissen, Bedell, Spina2018, Liu2020, elemage4_Nissen2020}. Similarly, we attempted to correct for it by following the methodology of \cite{Spina2016_elemage3} and \cite{elemage4_Nissen2020}. We derived a best linear fit to each element as a function of stellar age using the 20 TPHS samples for each PHS. We did not include PHSs when deriving the linear fit, because of the possibility of stellar-abundance variations associated with the existence of planets. The measured slope ($m$) and intercept ($b$) of each element were applied to correct for GCE effects in the following manner:

\begin{equation} \label{GCE}
	[\text{X}/\text{Fe}]_{\text{correct}} = [\text{X}/\text{Fe}]_{\text{raw}} - (m \times \text{age} + b),
\end{equation}
\noindent where X represents each element.

Note that, unlike the previous studies, we did not attempt to homogeneously apply the same GCE correcting parameter ($m$, $b$) to all samples. Rather, we calculated the GCE correction parameter for individual PHSs separately, using the TPHSs selected for each PHS, and then applied the slope and intercept to correct the GCE for each PHS. We believe that our GCE correction more clearly reflects the evolution effects of PHS by using TPHSs, which have similar atmospheric parameters with PHS within various age ranges. 

For the Sun, the values of the GCE correction parameters did not change significantly by varying the choice of the selected TPHS number from 10 to 30. We also found that the Sun's abundance trend against \tc\ was nearly identical regardless of the selected number of TPHSs. For the PHS sample, however, we found that the abundances of Al and Mn after GCE correction change significantly when we decrease the TPHS number from 20 to 10, but not when it was increased from 20 to 30 TPHSs for each PHS. This is the main reason for choosing 20 TPHS. We corrected the GCE effect for each TPHS in the same way as well, using the selected TPHSs of each PHS.

\begin{figure}
		\includegraphics[width=\linewidth]{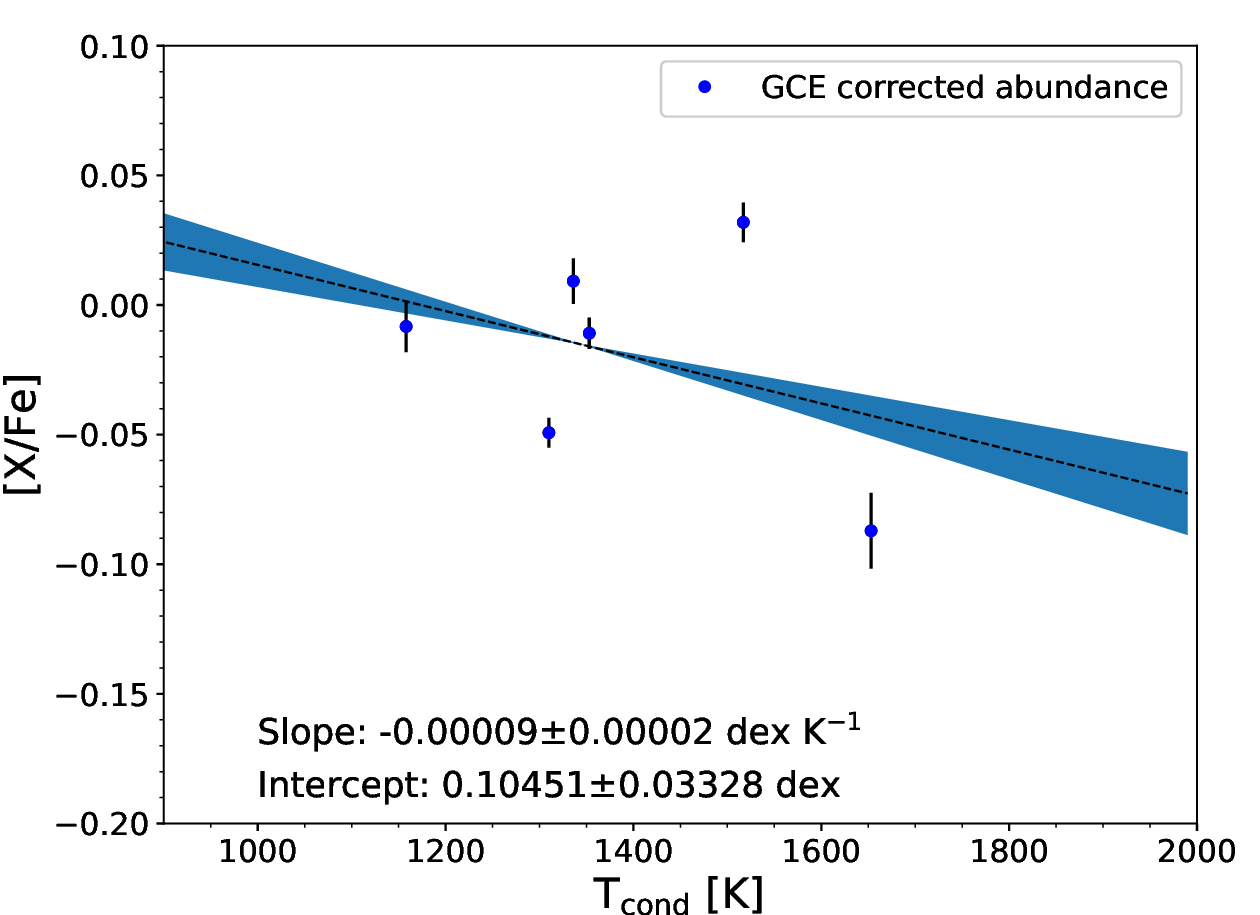}
		\caption{The Sun's GCE-corrected abundance trend as a function of \tc. Each dot represents the six elements (Mn, Si, Mg, Ni, Ca, and Al in order of \tc) placed at corresponding \tc, with error bars indicating the 1$\sigma$ of abundance value measured by 1000 random resamplings of the TPHSs, while correcting for GCE effects. The black-dotted line is the best-fit line, with the blue-shaded region displaying the 1$\sigma$ of the slopes. The derived slope, intercept, and their 1$\sigma$ values are indicated in the legend.}

	\label{fig:Fig1}

\end{figure}

\begin{figure*}
	\begin{center}
		\includegraphics[width=\linewidth]{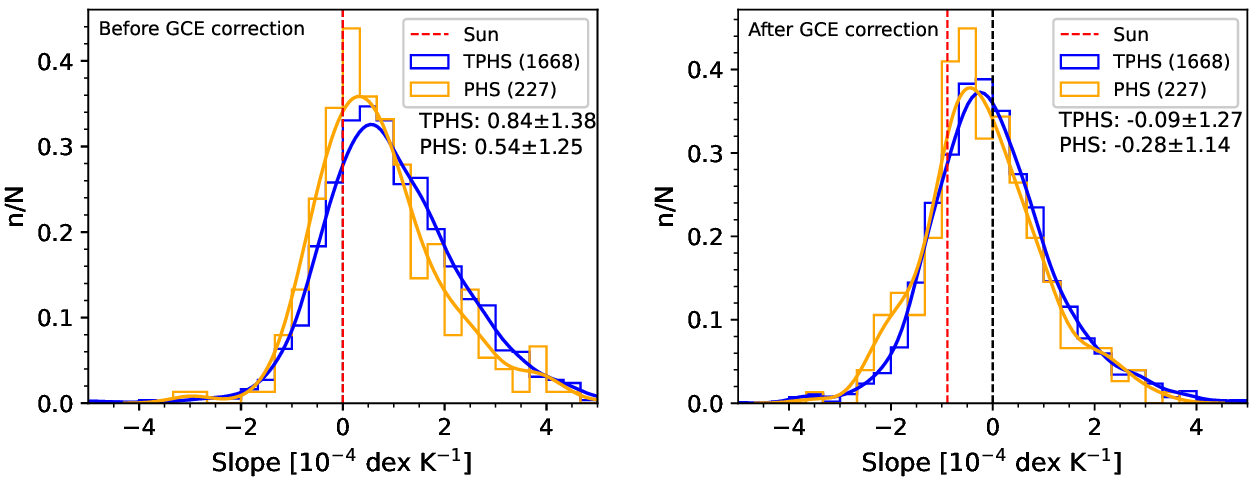}
		\caption{Distributions of the abundance-\tc\ slopes for our PHSs (orange) and TPHSs (blue). The left panel represents the results before the GCE correction, while the right panel is after the GCE correction. The smoothed distribution is a kernel density estimation (KDE). The median slope and 1$\sigma$ value for each group are denoted in each panel. The red and black dashed vertical lines are the Solar slope and zero point, respectively. The distribution is normalized to the total number of stars in each group.}
        \label{fig:Fig2}
	\end{center}
\end{figure*}

\vskip 1cm
\section{Derivation of the Abundance-\tc\ Slope} \label{sec:4}

We employed the method of the abundance gradient over \tc, which is the most widely used one to identify the chemical depletion or enhancement in a star. We have learned from previous studies that the accuracy of the chemical abundances of the PHSs plays a pivotal role in drawing a robust abundance slope over \tc. At the same time, it is imperative that the comparison sample (TPHSs in our case) accurately reflects the evolutionary stage of their counterparts as closely as possible. To robustly detect the variation of the abundance ratios, we made use of high-\tc\ refractory elements with the most reliable abundance determinations, as reported in \cite{APG16}. These are Mg, Al, Si, Ca, Mn, and Ni. We did not include volatile elements with \tc\ under 900 K, because they fall outside the steepest portion of the depletion trend \citep{Mel, Cham_TEABUND, Bedell}.

Figure \ref{fig:Fig1} shows an example of the derived abundance trend as a function of \tc. The figure shows the abundance slope of refractory elements of the Sun after the GCE correction. Each dot represents the six elements (Mn, Si, Mg, Ni, Ca, and Al in order of \tc) placed at their corresponding \tc; error bars were calculated from 1000 random resamplings of the Sun's twins while correcting for GCE. The dotted line is the abundance gradient derived from linear fit to the 6 elemental abundances. The blue-shaded region shows the 1$\sigma$ variation of the slope values, and the figure clearly indicates that the Sun is deficient with the refractory elements after the GCE correction.

Other studies also reported the depletion of refractory elements in the Sun. For instance, \cite{Bedell} analyzed 79 Solar twins to find evidence of the depletion of 26 elements in the Sun after the GCE correction. Similarly, \cite{Liu2020} demonstrated that the Sun's refractory elements are relatively depleted compared to its analog stars. \cite{Niba} successfully reproduced the results of \cite{Bedell} using only five refractory elements from APOGEE DR16 \citep{APG16}, supporting the depletion argument.

Using six refractory elements from APOGEE DR17, we reach the same conclusion. Note that our approach is somewhat different from the previous studies, in that our abundance slope is not derived from the star-to-star abundance differences between the PHSs and NPHSs, as done in most of the other studies \citep{Mel, Bedell, Liu2020, Niba, Tau}. Instead, the derived slope value represents the trends of the Sun's abundance values after correction of the chemical evolution of the MW.

\vskip 1.5cm
\section{Results} \label{sec:5}

\subsection{Distribution of Abundance-\tc\ Slopes} \label{subsec:5.1}

The primary reasons for the depletion of the Sun's refractory elements are still under 
discussion. \cite{Mel} argued that this depletion arose due to the presence of rocky planets such as Earth. However, other studies \citep{Gon2013_nosignal, Liu2020, Mishenina_2021} failed to find evidence to support this hypothesis, even with much higher-quality abundance analyses for limited samples. Extending the idea to other planetary systems, some studies reported a weak correlation between planetary mass and stellar chemical abundance \citep{Mishenina_2021, Tau}, while others using large survey data concluded that there is no clear sign of the correlation resulting from the existence of planets \citep{Niba, Behmard_2023_APG17}.

Based on our much larger sample of PHSs and TPHSs, we have investigated these issues by comparing the abundance-\tc\ slopes of PHSs with those of the TPHSs (stars without planetary detections), as shown in Figure \ref{fig:Fig2}. The abundance-\tc\ slopes of TPHSs are measured in the same manner as for the PHSs, using 20 TPHSs for each TPHS. In the figure, the orange lines represent the distribution of the PHSs and the blue lines are for the TPHSs. The left panel is before the GCE correction, while the right panel is after the GCE correction. 

Comparison of the two panels in Figure \ref{fig:Fig2} reveals that the GCE-corrected samples exhibit a consistently greater fraction of negative slopes, indicating that the GCE correction is an important step to apply when examining any abundance depletion or enhancement. Inspection of the right panel suggests that the PHSs do not exhibit a significant difference in the distribution of slopes from the TPHSs, although we observe a small enhancement of relative numbers of stars around the slope of --2.0$\times10^{-4}$ dex\,K$^{-1}$ for the PHSs, which was not observed before the GCE correction. We conclude that the existence of planetary systems does not strongly influence the chemical properties of their host stars, as pointed out by previous studies. To quantitatively evaluate the significance of the difference in the distributions, we carried out a Kolmogorov–Smirnov (KS) two-sample test under the null hypothesis that the two groups of stars share the same parent chemical properties. From the distribution with the GCE correction (right panel), we obtained a $p$-value of 0.1, implying that we cannot rule out the null hypothesis completely. We have thus focused on the existence of any chemical discrepancies among the PHSs with different planet populations and architectures.

\begin{figure}[!t]
		\includegraphics[width=\linewidth]{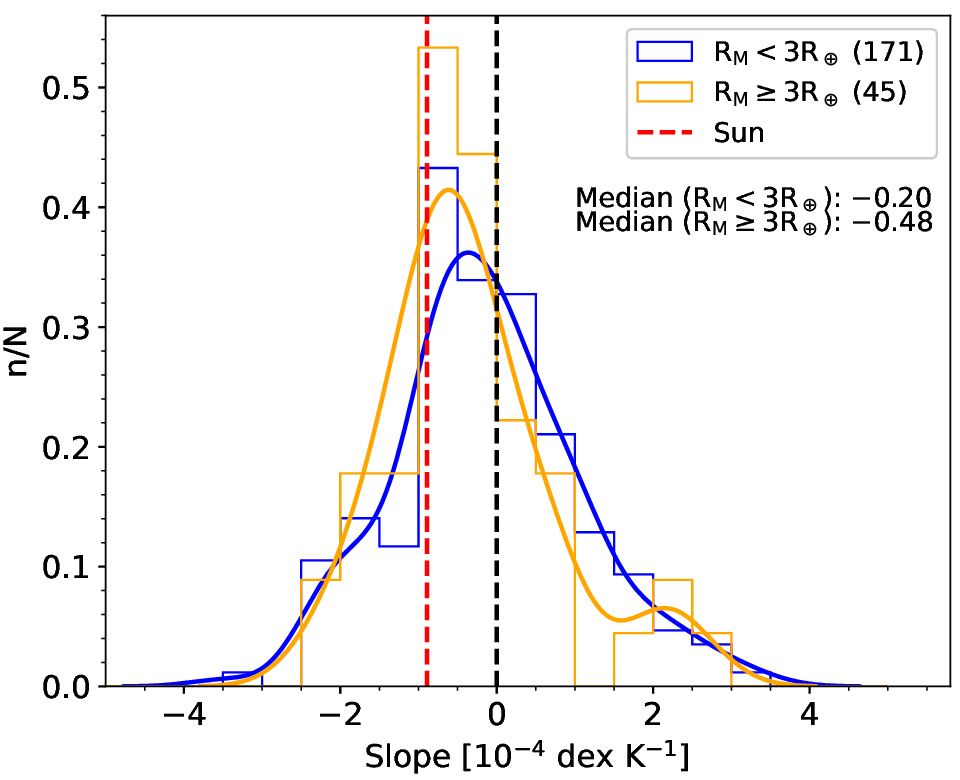}
		\caption{Distributions of the abundance-\tc\ slopes for two sub-groups of stars, after application for GCE effects, divided by the mean radius (\rmean) of the planetary system for each PHS. The blue histogram represents for PHSs with \rmean\ $<$ 3 \re, while the orange one applies to PHSs with \rmean\ $\geq$ 3 \re. See Section \ref{subsec:5.2} for details of the choice of 3 \re\ as the dividing line. The smoothed distribution is a KDE. The median slope value for each sub-group is denoted in the legend. The red-dashed vertical line is the Solar slope. The distributions are normalized by the total number of stars in each 
        sub-group.}
        \label{fig:Fig3}
\end{figure}

\subsection{Terrestrial Versus Jovian Planets} \label{subsec:5.2}

Before exploring how the different types of planets affect the chemical characteristics of their host stars, we need to distinguish the terrestrial planets from the Jovian planets by considering their observables such as mass and radius. Most recent studies, such as \cite{Mishenina_2021} and \cite{Tau}, attempted to correlate the planetary system properties with the chemical composition of their host stars by the use of the planetary mass. Out of the 338 planets in our sample, only 80 have known mass information, while the radius information is available for 323 planets, determined by transit observations. Since more planets provide more robust statistics, we decided to use the planetary-radius information to distinguish the terrestrial from the Jovian planets, and examine any connection to the abundance patterns of their host stars. We used a  total of 216 PHSs with known planets and their radii.

No super-Earth located at larger than 3 \re\ has been found to date, and according to the planetary orbital period-radius diagram, the Neptune desert spans from 2 to 10 \re\ with an orbital period of less than 10 days. Referring to these two observational constraints, we decided to set the distinguishing criterion at 3 \re\ between rocky and gas giant planets. If the mean radius (\rmean) of all known planets in a planetary system is less than 3 \re, we assume that the system is dominated by terrestrial planets, and by gas giants for \rmean\ $\geq$ 3 \re. It should be noted that our dividing condition for the types of planets is not an absolute standard, and is subject to change as more diverse planetary systems are discovered in different environments. Our sample comprises 155 single planetary systems and 61 multiple planetary systems.

To gain a deeper understanding of how the formation of different types of planets (e.g., rocky versus gaseous) influences the chemistry of their host stars, once again we derived the abundance slope of the refractory elements over \tc\ for the sub-groups of PHSs with terrestrial or Jovian planetary systems. Figure \ref{fig:Fig3} presents these distributions. The blue histogram represents the terrestrial planets with \rmean\ $<$ 3 \re, while the orange histogram is for the Jovian planets with \rmean\ $\geq$ 3 \re. The red-dashed vertical line is the Solar slope.

\begin{figure}[!t]
	\begin{center}
            \includegraphics[width=0.94\linewidth]{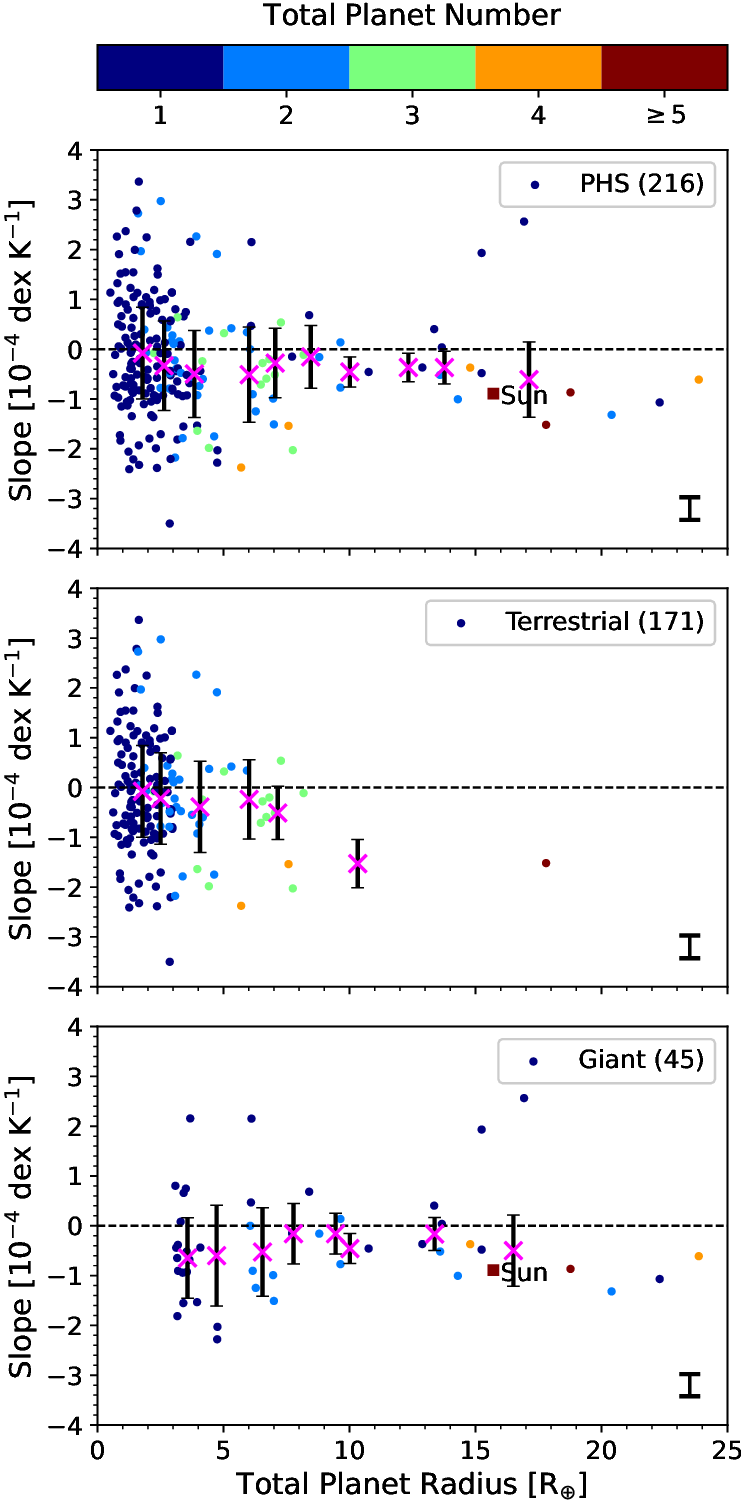}
		\caption{Elemental-abundance slopes as a function of the total radius of a planetary system; $R_{\rm tot}$ is the sum of the radii of all planets that belong to a star. Our full PHS sample (top panel) is divided into terrestrial (middle panel) and giant planets (bottom panel) by the mean radius of 3 $R_{\oplus}$ of each planetary system. The color code indicates the total number of planets in each PHS. The magenta symbols represent the median values of slopes grouped by 3 $R_{\oplus}$. Each bin is overlapped with the neighboring bin by 1.5 $R_{\oplus}$, and the error bar is the standard deviation in each bin. Note that the right-most bin includes all points larger than the bin value. The square in the top and bottom panels denote the Sun's slope. The typical error bar of each slope is indicated in the lower-right corner in each panel.} 
		\label{fig:Fig4}
	\end{center}
\end{figure}

\begin{figure*}[!t]
	\begin{center}
		\includegraphics[width=\linewidth]{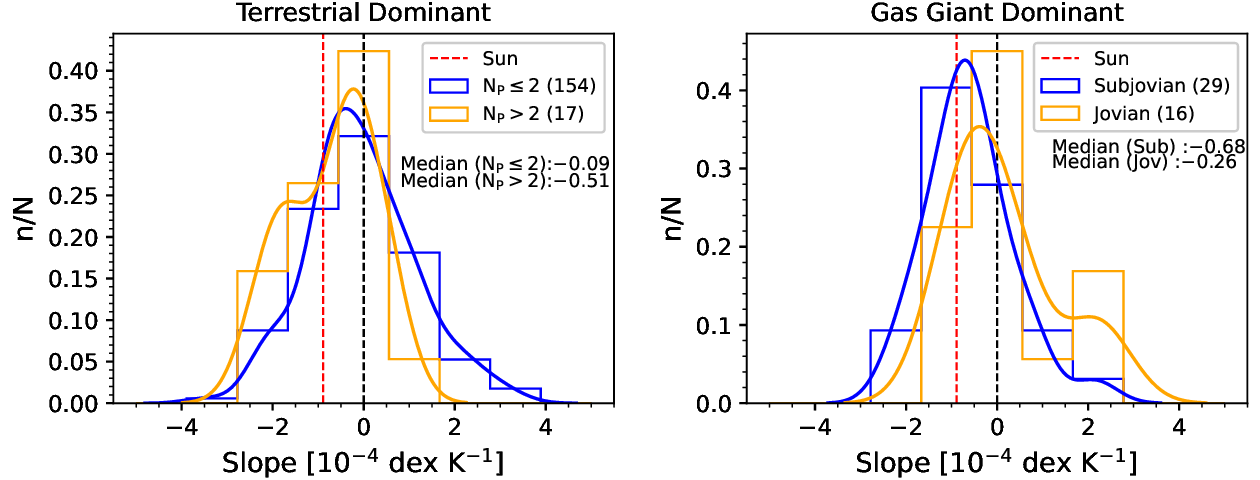}
		\caption{Left panel: distribution of the abundance slopes for the selected terrestrial planets. The planetary systems are further divided into two 
        sub-groups: one with one or two planets (blue lines) and the other with more than two planets (orange lines). The distributions are normalized by the total number of PHSs in each category. Curved lines represent the KDE of each sub-group. The median slope for each sub-group is listed in the legend. The PHSs with larger numbers of planets are shifted toward more negative slopes. Right panel: same as in the left panel, but for the sub-Jovian (blue lines) and Jovian (orange lines) sub-groups. We considered the systems with \rmean\ $<$ $5~R_{\oplus}$ as sub-Jovian, while the one with \rmean\ $\geq 5~R_{\oplus}$ as Jovian.}
	\label{fig:Fig5}
	\end{center}
\end{figure*}

One interesting aspect of Figure \ref{fig:Fig3} is that the peak and overall shape of the larger mean-radius sub-group is shifted to more negative slope values compared with those of the smaller mean-radius sub-group. The median value of the smaller-radius sub-group is also twice as high as the larger-radius sub-group. This implies that the sub-group dominated by giant planets exhibits slightly more chemical depletion than that dominated by terrestrial planets. Intriguingly, the peak of the Jovian sub-group distribution is closer to the Solar slope than that of the terrestrial sub-group. Note that, according to our definition, our Solar System falls into the category of the Jovian-dominated sub-group.

To further examine possible connections of the planet number and type with the chemical characteristics of their host stars, we consider the slope distributions as a function of the total radius ($R_{\rm tot}$) of the planets for each PHS, as shown in Figure \ref{fig:Fig4}. $R_{\rm tot}$ is the sum of the radii of all planets that belong to a star. The top panel is the full PHS sample, while the middle and bottom panels are for the systems dominated by terrestrial and giant planets, respectively, following the same division as in Figure \ref{fig:Fig3}. The color code represents the number of planets in each system. The magenta symbols are the median value of the slopes in a bin of 3 $R_{\oplus}$ overlapped 50\% with the neighboring bin, and the error bar of each PHS is the standard deviation in each bin. The typical error bar of each slope is denoted at the bottom right corner of each panel. As the Sun belongs to the Jovian-dominant sub-group in our definition, the Solar slope appears in the bottom panel.

Inspection of Figure \ref{fig:Fig4} reveals the following features. First, we see that most of the PHSs are dominated by a single-planet system. The top plot exhibits a tendency for having more negative slopes for $R_{\rm tot} > 5$ \re, with more planets, indicating a greater depletion of the refractory elements. These trends can be read off from the middle and bottom panels. The middle panel implies that, the more rocky planets a PHS has, the more depletion of the refractory elements has taken place. In other words, the depletion trend becomes stronger as the total radius and the number of planets increase. The bottom plot also shows a small, but relatively consistent negative slope, and this tendency becomes even stronger for $R_{\rm tot} > 15$ \re\ and for more planets. However, the sample size of the giant sub-group is much smaller than that of the terrestrial sub-group, which may suffer from small number statistics. We discuss in depth the significance and implications of these findings in section \ref{subsec:6.3}.

\begin{figure*} [!t]
	\begin{center}
		\includegraphics[width=\linewidth]{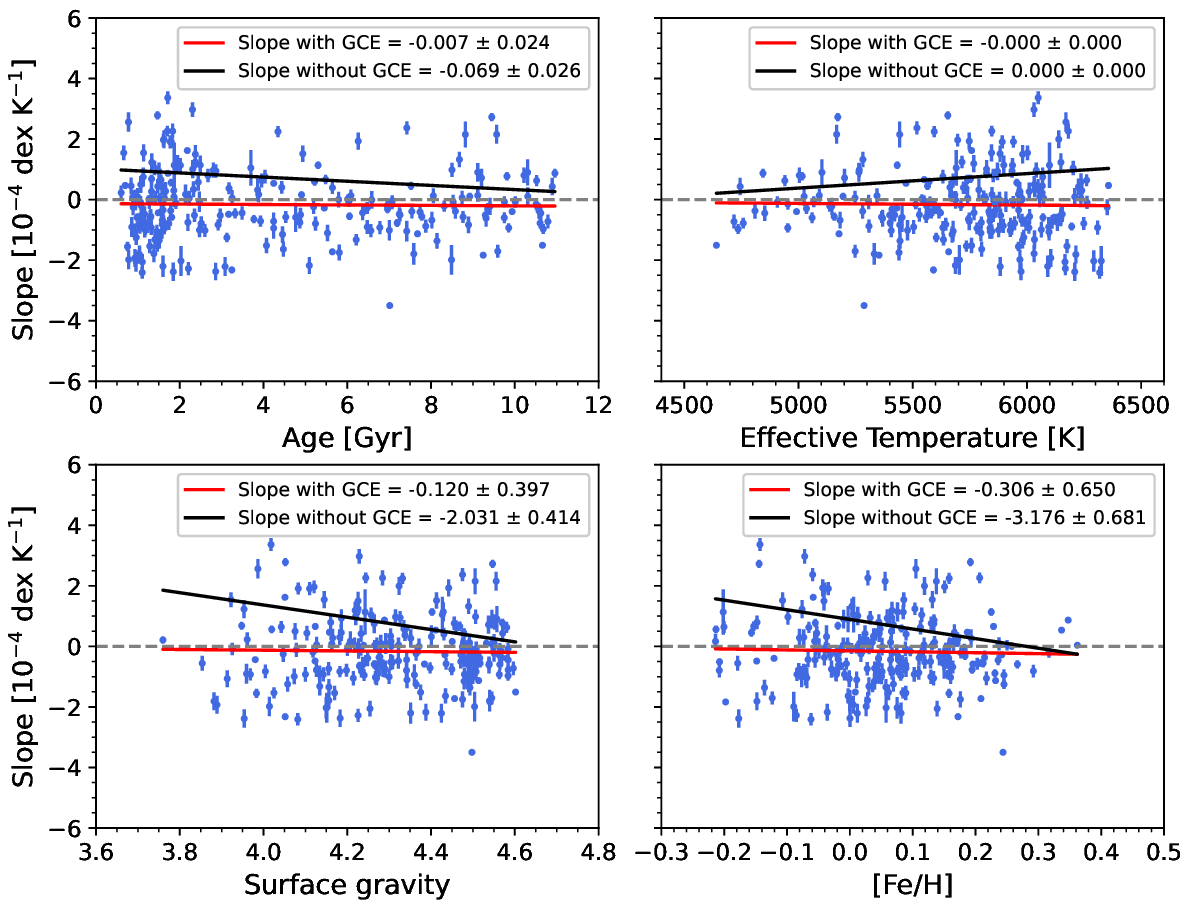}
		\caption{Abundance-\tc\ slopes of the PHS sample, corrected for GCE effects, as functions of the stellar parameters: age (upper left), effective temperature (upper right), surface gravity (lower left), and metallicity (lower right). The error bar on each star is calculated from 1000 random resamplings of the TPHSs for each PHS, when correcting for GCE. The red line in each panel is the linear fit to the distribution, while the black line comes from the PHSs without the GCE correction. The derived slope and its error are denoted at the top of each panel.}
	\label{fig:Fig6}
	\end{center}
\end{figure*}

\subsection{Number of Planets and their Type} \label{subsec:5.3}

Figures \ref{fig:Fig3} and \ref{fig:Fig4} indicate that the number and type of planets may be related to the general chemical properties of their host stars. We investigate this possibility more closely in Figure \ref{fig:Fig5}. We divided the terrestrial group into two sub-groups by the number of planets (\np) and the gas-giant group according to their type. The left panel of Figure \ref{fig:Fig5} represents the terrestrial group. The blue histogram is for \np\ $\leq$ 2, while the orange histogram is for \np\ $>$ 2. The smoothed curves are the KDE of each histogram. We notice that there is a tendency for the host stars having more rocky planets to be more deficient in their refractory elements, as they exhibit more negative slopes. The median value of the slopes of the systems with \np\ $\leq$ 2 is five times smaller than the systems with \np\ $>$ 2. Overall, the fraction of PHSs with the negative slopes is 56.8$\%$ for \np\ $\leq$ 2, but 85.7$\%$ for \np\ $>$ 2. These fractions indicate a greater depletion of the refractory elements for a star with more rocky planets.

The right panel of Figure \ref{fig:Fig5} shows the slope distribution of the PHSs with the sub-Jovian (blue lines) and Jovian planets (orange lines). The division line for these two sub-groups is the mean radius of 5 \re: sub-Jovian for  \rmean\ $<$ 5 \re, Jovian for \rmean\ $\geq$ 5 \re. The plot clearly illustrates that the peaks between the two sub-groups and the overall fraction of the PHSs with negative slopes differ significantly. The sub-Jovian sub-group has more negative slopes, demonstrating a stronger depletion of the refractory elements than for the Jovian sub-group. This suggests that the PHSs with smaller gas giants are expected to be more depleted in the refractory elements than the ones with the larger gas giants. We obtained consistent results even if we decreased the mean planet radius to \rmean\ = 2 $R_{\oplus}$ to divide the sub-groups, which is frequently used as a threshold for dividing the terrestrial planets. Our conclusion from Figures \ref{fig:Fig4} and \ref{fig:Fig5} is that the number and type of the planets can influence the chemical properties of their host stars. However, we also note that the sample size of the sub-groups is not sufficiently large provide a strong conclusion (See section \ref{subsec:6.3}).

\vskip 2.5cm
\section{Discussion} \label{sec:6}

\subsection{Correlation between Stellar Parameters and Abundance-\tc\ Slopes} \label{subsec:6.1}

It remains possible that the different behaviors seen in the abundance-\tc\ slopes between the terrestrial and giant planets may be caused by different spectral and luminosity classes of their host stars, as well as their age and metallicity. Previous studies were also concerned whether or not the magnitude and the sign of the abundance gradient with \tc\ depend on stellar parameters. For instance, \cite{Rami_2009} reported that the abundance gradients with \tc\ measured by the refractory elements with \tc\ $>$ 900 K exhibited correlations with surface gravity and metallicity, but not with effective temperature. In addition, \cite{Adi2014} demonstrated that the abundance trend, regardless of the presence of the planet, had a strongly negative correlation with surface gravity and a positive correlation with stellar age. On the other hand, \cite{Gon2013_nosignal}, using 25 elements, starting from carbon with \tc\ = 40 K, could not find any correlation of their abundance slopes with the stellar temperature and metallicity. Similarly, \cite{Liu2020} claimed that there is no significant correlation between the abundance slopes of PHSs and stellar age, metallicity, or temperature with a 3$\sigma$ confidence, while the NPHSs have a large scatter with the three parameters.

The present lack of consensus requires us to carefully examine possible correlations between the stellar parameters and the abundance slope in our program stars.  Figure \ref{fig:Fig6} shows the derived slopes of the PHSs as functions of age (upper left), effective temperature (upper right), surface gravity (lower left), and metallicity (lower right). The red line in each panel is the linear fit to the PHSs distribution after the GCE correction, while the black line is the fit to these stars without the GCE correction. The error bar for each PHS is calculated from 1000 random resamplings of TPHSs, as in Figure \ref{fig:Fig1}. These plots clearly exhibit no significant trends of the abundance slopes against the various stellar parameters, after applying the GCE correction. This figure demonstrates that the disparate patterns found in the abundance slopes for the stars dominated by rocky and gas giant planets do not arise from different evolution stages, chemistry, or age of the host stars.

However, the abundance-\tc\ slopes of our uncorrected PHSs do exhibit negative correlations with respect to surface gravity and metallicity, and a positive correlation with effective temperature, similar to other studies that did not correct for GCE \cite[e.g.,][]{Rami_2009, Adi2014}. This once more highlights the importance of the correction of the GCE when it comes to deriving the abundance slopes. The main reason for the lack of the correlation between the abundance-\tc\ slope and stellar parameters may stem from the fact that, unlike previous papers applying the same GCE correction to all the samples, we selected the TPHSs sharing the same evolutionary states with each PHS to correct for the effects of GCE. It appears that our approach has removed any clear correlations with stellar parameters.

\subsection{Correlation of Chemical Depletion with the Presence of Planets} \label{subsec:6.2}

We have found that the Sun exhibits a negative abundance-\tc\ slope of refractory elements after correcting for GCE, using twin stars in similar evolutionary states. This implies that some of the refractory elements are depleted, as reported in other previous studies \cite[e.g.,][]{Mel, Bedell}. 

Our PHSs indicate a slightly lower abundance of refractory elements compared to their counterparts of TPHSs, although most of the abundance-\tc\ slope values of PHSs agree well with the previous results of \citet{Mel, Bedell, Niba}, on the order of 4$\times~10^{-4}$ dex K$^{-1}$. However, we recall that a K-S two-sample test cannot reject the null hypothesis that the PHS and TPHSs abundance slopes are drawn from the same population, as shown in Figure \ref{fig:Fig2}. This argument is in line with other studies claiming very weak or no correlation between the presence of the planets and chemical anomalies of their host stars.

As illustrated in Figures \ref{fig:Fig3}, \ref{fig:Fig4}, and \ref{fig:Fig5}, all terrestrial PHSs do not show refractory-element depletion; rather, some of them exhibit an enhancement of refractory elements. Some studies \citep[e.g.,][]{Liu2020} attempted to explain the enhancement of the heavy elements as due to the engulfment of planets. That is, planetary bodies could be engulfed by their host star during its red giant phase, contributing to the higher abundance trends \citep{Pinso_engulf, Saffe_engulf, engulf_common,engulf_Urich}. However, our sample of PHSs selected from APOGEE DR17 primarily consist of dwarf stars and sub-giants whose evolutionary stage is too early to engulf their planets. The case of the positive slopes (or the abundance enhancements) requires more investigation.

We find that our PHSs have two unique chemical features associated with their planet populations: (1) The more rocky planets a star has, the greater the depletion of refractory elements in its host, and (2) the greater the total number of planets a star has, the more depletion of refractory elements in its host (see Figure \ref{fig:Fig4}). Only a few studies have attempted to explore these aspects, probably due to the limited sample size of the high-resolution spectroscopic data \citep{Rami_2009, Gon2013_nosignal, Niss2015, Bedell, Mishenina_2021} and the different sample-selection criteria with limited planet information \citep{Liu2020, Tau}. For instance, by analyzing 24 elements from 740 bright slow-rotating stars, including 25 PHSs, \cite{Tau} claimed that the metallicity of the dwarf stars hosting low-mass planets is usually high, while those with high-mass planets exhibit a greater diversity. Still, to clarify the chemical effects of the planet on the host star, additional studies with larger numbers of samples are needed.

\subsection{The Impact of Planets on Chemical Anomalies of their Host Stars} \label{subsec:6.3}

As described above, the depletion level of the refractory elements for our PHSs is correlated with their planet population. This provides us with an opportunity to relate their behavior to the suggested planet-formation scenarios. Following the rocky-planet argument of \cite{Mel}, 60 \% of our samples with negative abundance slope should be depleted with refractory elements, implying the presence of rocky planets. However, because all of our PHSs do not possess rocky planets, but some of them also include giant planets, the terrestrial planets alone cannot explain the abundance-\tc\ slope distribution; we need the giant planets to explain the abundance-slope distributions as well. \cite{Booth_engulf_gap} showed that the pressure gap generated by forming giant planets can prevent refractory material from accreting onto their host stars. This scenario may explain the negative slope of the PHSs with large total-radius planets. It follows that our findings are likely to be explained by the rocky-planet theory of \cite{Mel} and the giant-planet hypothesis by \cite{Booth_engulf_gap}.

Figure \ref{fig:Fig5} shows a very interesting aspect of this problem, that the sub-Jovian population exhibits more negative abundance slopes than that of the Jovian population. According to the work by \cite{Planet_formation_efficiency}, the opening gap of a protoplanetary disk from planetary formation hinders the accretion of heavy elements to the host star, leaving room for an example of a wide-binary model that explains the difference in the element-abundance ratios of each star, depending on the formation position of its giant gas planets. This model may explain the reason for the lower median abundance slope of the giant planets compared with that of the terrestrial planets seen in Figures \ref{fig:Fig3} and \ref{fig:Fig5}. Thus, this may raise various questions about the planetary-formation process of Neptune-like and Jupiter-like planets.

Our present findings cannot be solely explained by current planet-formation theories. The difference in the abundances of the refractory elements between PHSs and TPHSs and the wide range of the abundance slopes may require us to take into account a variety of planet-induced effects as well. \cite{Liu2020} pointed to planet-induced effects, such as the sequestration of rocky material (depleted with refractory elements) and the engulfment of planets (enhanced with refractory elements) \citep{Pinso_engulf, Saffe_engulf, Booth_engulf_gap} for the diverse abundance trends in the PHSs. These are related to the timescale, accretion efficiency of matter, and occurrence rate of the planet formation or protoplanetary-disk chemical evolution \citep{planet_formation_Johansen, Planet_formation_efficiency}. These effects could potentially affect the chemical composition of the PHSs seen in this study.

One of the planet-induced effects to consider to explain the diverse abundance-\tc\ slopes among our PHSs is the planet-occurrence rate. Even though it is actively debated at present \citep{Hsu}, it has been reported to be as high as 1.0 \citep{Zink}. Microlensing studies \citep[e.g.,][]{Cassan} also report a high probability for a given star to have a planet. These observational results suggest that most stars host planets that have not yet been discovered. Detecting the presence of planets by transit studies makes it difficult to discover long-period planets due to their weak signals and observational limitations. Consequently, we cannot rule out that our selected PHSs may contain more undetected planets. In turn, a high planet-occurrence rate greatly complicates the identification of abundance differences among PHSs with different types of stars. Although it is unclear how much the existence of undiscovered planets influences our analysis, it is certain that, depending on the type and number of undetected planets among our PHSs, our results are subject to change.

\vskip 1cm
\section{Summary and Conclusions} \label{sec:7}

We have analyzed a sample of PHSs selected from APOGEE DR17 to search for chemical differences compared to the TPHSs without planets. We found very weak evidence that the PHSs are more depleted with refractory elements than the TPHSs. However, we demonstrated that the amount of material contributed to planetary formation can result in the depletion of refractory elements in their host stars. Overall, the PHSs with giant planets are more depleted with heavy elements than the ones with rocky planets. Among the PHSs with rocky planets, the more rocky planets a star has, the more depletion of the refractory elements a host star exhibits. Interestingly, the sub-Jovian hosting stars exhibit more chemical depletion than the Jovian hosting stars. This may raise various questions about the planetary-formation process of Neptune-like and Jupiter-like planets.

Our study indicates that the wide range of observed abundance-\tc\ slopes for PHSs cannot be explained solely by present planet-formation scenarios; we also need to understand how planet-induced effects affect the characterization of the PHSs. Our understanding of the relationship between the elemental-abundance patterns of host stars and the formation of planets is still far from complete, and many complexities remain to be unraveled. Observations with the James Webb Space Telescope \citep{jwst} should help clarify the relationship(s) between planet formation and the chemistry of host stars, with its capability to stack enormous direct infrared images of planetary systems, and test protoplanetary disk-formation theories around host stars in various stages of evolution.

\vspace{10mm}
\addcontentsline{toc}{section}{acknowledgments}

Y.S.L. acknowledges support from the National Research Foundation (NRF) of Korea grant funded by the Ministry of Science and ICT (NRF-2021R1A2C1008679). Y.S.L. also gratefully acknowledges partial support for his visit to the University of Notre Dame from OISE-1927130: The International Research Network for Nuclear Astrophysics (IReNA), awarded by the US National Science Foundation. T.C.B. acknowledges partial support for this work from grant PHY 14-30152; Physics Frontier Center/JINA Center for the Evolution of the Elements (JINA-CEE), and OISE-1927130: The International Research Network for Nuclear Astrophysics (IReNA), awarded by the US National Science Foundation. D.L. acknowledges support from the National Research Foundation of Korea to the Center for Galaxy Evolution Research (2022R1A6A1A03053472). Y.K.K. acknowledges support from Basic Science Research Program through the NRF of Korea funded by the Ministry of Education (NRF-2021R1A6A3A01086446).

Funding for the Sloan Digital Sky Survey IV has been provided by the Alfred P. Sloan Foundation, the U.S. Department of Energy Office of Science, and the Participating Institutions. 

SDSS-IV acknowledges support and resources from the Center for High Performance Computing at the University of Utah. The SDSS website is \url{www.sdss4.org}. SDSS-IV is managed by the Astrophysical Research Consortium for the Participating Institutions of the SDSS Collaboration including the Brazilian Participation Group, the Carnegie Institution for Science, Carnegie Mellon University, Center for Astrophysics | Harvard \& Smithsonian, the Chilean Participation Group, the French Participation Group, Instituto de Astrof\'isica de Canarias, The Johns Hopkins University, Kavli Institute for the Physics and Mathematics of the Universe (IPMU) / University of Tokyo, the Korean Participation Group, Lawrence Berkeley National Laboratory, Leibniz Institut f\"ur Astrophysik Potsdam (AIP), Max-Planck-Institut f\"ur Astronomie (MPIA Heidelberg), Max-Planck-Institut f\"ur Astrophysik (MPA Garching), Max-Planck-Institut f\"ur Extraterrestrische Physik (MPE), National Astronomical Observatories of China, New Mexico State University, New York University, University of Notre Dame, Observat\'ario Nacional / MCTI, The Ohio State University, Pennsylvania State University, Shanghai Astronomical Observatory, United Kingdom Participation Group, Universidad Nacional Aut\'onoma de M\'exico, University of Arizona, University of Colorado Boulder, University of Oxford, University of Portsmouth, University of Utah, University of Virginia, University of Washington, University of Wisconsin, Vanderbilt University, and Yale University.

This research has made use of the NASA Exoplanet Archive, which is operated by the California Institute of Technology, under contract with the National Aeronautics and Space Administration under the Exoplanet Exploration Program.

This work presents results from the European Space Agency (ESA) space mission Gaia. Gaia data are being processed by the Gaia Data Processing and Analysis Consortium (DPAC). Funding for the DPAC is provided by national institutions, in particular the institutions participating in the Gaia MultiLateral Agreement (MLA). The Gaia mission website is \url{https://www.cosmos.esa.int/gaia}. The Gaia archive website is \url{https://archives.esac.esa.int/gaia.}

\facilities{Ir\'en\'ee du Pont telescope (APOGEE-2), Sloan 2.5m telescope (APOGEE), Gaia,
NASA Exoplanet Archive.}

\software{numpy \citep{numpy}, matplotlib \citep{matplotlib}, pandas \citep{pandas}, scipy \citep{2020SciPy-NMeth}, scikit-learn \citep{scikit-learn}, astropy \citep{astropy}.}

\bibliography{ref}

\begin{thebibliography}{}
\expandafter\ifx\csname natexlab\endcsname\relax\def\natexlab#1{#1}\fi
\providecommand{\url}[1]{\href{#1}{#1}}
\providecommand{\dodoi}[1]{doi:~\href{http://doi.org/#1}{\nolinkurl{#1}}}
\providecommand{\doeprint}[1]{\href{http://ascl.net/#1}{\nolinkurl{http://ascl.net/#1}}}
\providecommand{\doarXiv}[1]{\href{https://arxiv.org/abs/#1}{\nolinkurl{https://arxiv.org/abs/#1}}}

\bibitem[{{Abdurro'uf} {et~al.}(2022){Abdurro'uf}, {Accetta}, {Aerts}, {Silva Aguirre}, {Ahumada}, {Ajgaonkar}, {Filiz Ak}, {Alam}, {Allende Prieto}, {Almeida}, {Anders}, {Anderson}, {Andrews}, {Anguiano}, {Aquino-Ort{\'\i}z}, {Arag{\'o}n-Salamanca}, {Argudo-Fern{\'a}ndez}, {Ata}, {Aubert}, {Avila-Reese}, {Badenes}, {Barb{\'a}}, {Barger}, {Barrera-Ballesteros}, {Beaton}, {Beers}, {Belfiore}, {Bender}, {Bernardi}, {Bershady}, {Beutler}, {Bidin}, {Bird}, {Bizyaev}, {Blanc}, {Blanton}, {Boardman}, {Bolton}, {Boquien}, {Borissova}, {Bovy}, {Brandt}, {Brown}, {Brownstein}, {Brusa}, {Buchner}, {Bundy}, {Burchett}, {Bureau}, {Burgasser}, {Cabang}, {Campbell}, {Cappellari}, {Carlberg}, {Wanderley}, {Carrera}, {Cash}, {Chen}, {Chen}, {Cherinka}, {Chiappini}, {Choi}, {Chojnowski}, {Chung}, {Clerc}, {Cohen}, {Comerford}, {Comparat}, {da Costa}, {Covey}, {Crane}, {Cruz-Gon2011ez}, {Culhane}, {Cunha}, {Dai}, {Damke}, {Darling}, {Davidson}, {Davies}, {Dawson}, {De Lee}, {Diamond-Stanic}, {Cano-D{\'\i}az}, {S{\'a}nchez},
  {Donor}, {Duckworth}, {Dwelly}, {Eisenstein}, {Elsworth}, {Emsellem}, {Eracleous}, {Escoffier}, {Fan}, {Farr}, {Feng}, {Fern{\'a}ndez-Trincado}, {Feuillet}, {Filipp}, {Fillingham}, {Frinchaboy}, {Fromenteau}, {Galbany}, {Garc{\'\i}a}, {Garc{\'\i}a-Hern{\'a}ndez}, {Ge}, {Geisler}, {Gelfand}, {G{\'e}ron}, {Gibson}, {Goddy}, {Godoy-Rivera}, {Grabowski}, {Green}, {Greener}, {Grier}, {Griffith}, {Guo}, {Guy}, {Hadjara}, {Harding}, {Hasselquist}, {Hayes}, {Hearty}, {Hern{\'a}ndez}, {Hill}, {Hogg}, {Holtzman}, {Horta}, {Hsieh}, {Hsu}, {Hsu}, {Huber}, {Huertas-Company}, {Hutchinson}, {Hwang}, {Ibarra-Medel}, {Chitham}, {Ilha}, {Imig}, {Jaekle}, {Jayasinghe}, {Ji}, {Johnson}, {Jones}, {J{\"o}nsson}, {Katkov}, {Khalatyan}, {Kinemuchi}, {Kisku}, {Knapen}, {Kneib}, {Kollmeier}, {Kong}, {Kounkel}, {Kreckel}, {Krishnarao}, {Lacerna}, {Lane}, {Langgin}, {Lavender}, {Law}, {Lazarz}, {Leung}, {Leung}, {Lewis}, {Li}, {Li}, {Lian}, {Liang}, {Lin}, {Lin}, {Lin}, {Lintott}, {Long}, {Longa-Pe{\~n}a}, {L{\'o}pez-Cob{\'a}}, {Lu},
  {Lundgren}, {Luo}, {Mackereth}, {de la Macorra}, {Mahadevan}, {Majewski}, {Manchado}, {Mandeville}, {Maraston}, {Margalef-Bentabol}, {Masseron}, {Masters}, {Mathur}, {McDermid}, {Mckay}, {Merloni}, {Merrifield}, {Meszaros}, {Miglio}, {Di Mille}, {Minniti}, {Minsley}, {Monachesi}, {Moon}, {Mosser}, {Mulchaey}, {Muna}, {Mu{\~n}oz}, {Myers}, {Myers}, {Nadathur}, {Nair}, {Nandra}, {Neumann}, {Newman}, {Nidever}, {Nikakhtar}, {Nitschelm}, {O'Connell}, {Garma-Oehmichen}, {Luan Souza de Oliveira}, {Olney}, {Oravetz}, {Ortigoza-Urdaneta}, {Osorio}, {Otter}, {Pace}, {Padilla}, {Pan}, {Pan}, {Parikh}, {Parker}, {Peirani}, {Pe{\~n}a Ram{\'\i}rez}, {Penny}, {Percival}, {Perez-Fournon}, {Pinsonneault}, {Poidevin}, {Poovelil}, {Price-Whelan}, {B{\'a}rbara de Andrade Queiroz}, {Raddick}, {Ray}, {Rembold}, {Riddle}, {Riffel}, {Riffel}, {Rix}, {Robin}, {Rodr{\'\i}guez-Puebla}, {Roman-Lopes}, {Rom{\'a}n-Z{\'u}{\~n}iga}, {Rose}, {Ross}, {Rossi}, {Rubin}, {Salvato}, {S{\'a}nchez}, {S{\'a}nchez-Gallego}, {Sanderson}, {Santana
  Rojas}, {Sarceno}, {Sarmiento}, {Sayres}, {Sazonova}, {Schaefer}, {Schiavon}, {Schlegel}, {Schneider}, {Schultheis}, {Schwope}, {Serenelli}, {Serna}, {Shao}, {Shapiro}, {Sharma}, {Shen}, {Shetrone}, {Shu}, {Simon}, {Skrutskie}, {Smethurst}, {Smith}, {Sobeck}, {Spoo}, {Sprague}, {Stark}, {Stassun}, {Steinmetz}, {Stello}, {Stone-Martinez}, {Storchi-Bergmann}, {Stringfellow}, {Stutz}, {Su}, {Taghizadeh-Popp}, {Talbot}, {Tayar}, {Telles}, {Teske}, {Thakar}, {Theissen}, {Tkachenko}, {Thomas}, {Tojeiro}, {Hernandez Toledo}, {Troup}, {Trump}, {Trussler}, {Turner}, {Tuttle}, {Unda-Sanzana}, {V{\'a}zquez-Mata}, {Valentini}, {Valenzuela}, {Vargas-Gonz{\'a}lez}, {Vargas-Maga{\~n}a}, {Alfaro}, {Villanova}, {Vincenzo}, {Wake}, {Warfield}, {Washington}, {Weaver}, {Weijmans}, {Weinberg}, {Weiss}, {Westfall}, {Wild}, {Wilde}, {Wilson}, {Wilson}, {Wilson}, {Wolf}, {Wood-Vasey}, {Yan}, {Zamora}, {Zasowski}, {Zhang}, {Zhao}, {Zheng}, {Zheng}, \& {Zhu}}]{APG17}
{Abdurro'uf}, {Accetta}, K., {Aerts}, C., {et~al.} 2022, \apjs, 259, 35, \dodoi{10.3847/1538-4365/ac4414}

\bibitem[{{Adibekyan} {et~al.}(2014){Adibekyan}, {Gonz{\'a}lez Hern{\'a}ndez}, {Delgado Mena}, {Sousa}, {Santos}, {Israelian}, {Figueira}, \& {Bertran de Lis}}]{Adi2014}
{Adibekyan}, V.~Z., {Gonz{\'a}lez Hern{\'a}ndez}, J.~I., {Delgado Mena}, E., {et~al.} 2014, \aap, 564, L15, \dodoi{10.1051/0004-6361/201423435}

\bibitem[{{Adibekyan} {et~al.}(2012{\natexlab{a}}){Adibekyan}, {Sousa}, {Santos}, {Delgado Mena}, {Gonz{\'a}lez Hern{\'a}ndez}, {Israelian}, {Mayor}, \& {Khachatryan}}]{Adi2012}
{Adibekyan}, V.~Z., {Sousa}, S.~G., {Santos}, N.~C., {et~al.} 2012{\natexlab{a}}, \aap, 545, A32, \dodoi{10.1051/0004-6361/201219401}

\bibitem[{{Adibekyan} {et~al.}(2012{\natexlab{b}}){Adibekyan}, {Santos}, {Sousa}, {Israelian}, {Delgado Mena}, {Gonz{\'a}lez Hern{\'a}ndez}, {Mayor}, {Lovis}, \& {Udry}}]{Adibe_2012}
{Adibekyan}, V.~Z., {Santos}, N.~C., {Sousa}, S.~G., {et~al.} 2012{\natexlab{b}}, \aap, 543, A89, \dodoi{10.1051/0004-6361/201219564}

\bibitem[{{Astropy Collaboration} {et~al.}(2013){Astropy Collaboration}, {Robitaille}, {Tollerud}, {Greenfield}, {Droettboom}, {Bray}, {Aldcroft}, {Davis}, {Ginsburg}, {Price-Whelan}, {Kerzendorf}, {Conley}, {Crighton}, {Barbary}, {Muna}, {Ferguson}, {Grollier}, {Parikh}, {Nair}, {Unther}, {Deil}, {Woillez}, {Conseil}, {Kramer}, {Turner}, {Singer}, {Fox}, {Weaver}, {Zabalza}, {Edwards}, {Azalee Bostroem}, {Burke}, {Casey}, {Crawford}, {Dencheva}, {Ely}, {Jenness}, {Labrie}, {Lim}, {Pierfederici}, {Pontzen}, {Ptak}, {Refsdal}, {Servillat}, \& {Streicher}}]{astropy}
{Astropy Collaboration}, {Robitaille}, T.~P., {Tollerud}, E.~J., {et~al.} 2013, \aap, 558, A33, \dodoi{10.1051/0004-6361/201322068}

\bibitem[{{Bedell} {et~al.}(2014){Bedell}, {Mel{\'e}ndez}, {Bean}, {Ram{\'\i}rez}, {Leite}, \& {Asplund}}]{Bedell_precision}
{Bedell}, M., {Mel{\'e}ndez}, J., {Bean}, J.~L., {et~al.} 2014, \apj, 795, 23, \dodoi{10.1088/0004-637X/795/1/23}

\bibitem[{{Bedell} {et~al.}(2018){Bedell}, {Bean}, {Mel{\'e}ndez}, {Spina}, {Ram{\'\i}rez}, {Asplund}, {Alves-Brito}, {dos Santos}, {Dreizler}, {Yong}, {Monroe}, \& {Casagrande}}]{Bedell}
{Bedell}, M., {Bean}, J.~L., {Mel{\'e}ndez}, J., {et~al.} 2018, \apj, 865, 68, \dodoi{10.3847/1538-4357/aad908}

\bibitem[{{Behmard} {et~al.}(2023){Behmard}, {Ness}, {Cunningham}, \& {Bedell}}]{Behmard_2023_APG17}
{Behmard}, A., {Ness}, M.~K., {Cunningham}, E.~C., \& {Bedell}, M. 2023, \aj, 165, 178, \dodoi{10.3847/1538-3881/acc32a}

\bibitem[{{Belokurov} \& {Kravtsov}(2022)}]{accretion}
{Belokurov}, V., \& {Kravtsov}, A. 2022, \mnras, 514, 689, \dodoi{10.1093/mnras/stac1267}

\bibitem[{{Bennett} \& {Bovy}(2019)}]{solar_position_z}
{Bennett}, M., \& {Bovy}, J. 2019, \mnras, 482, 1417, \dodoi{10.1093/mnras/sty2813}

\bibitem[{{Berke} {et~al.}(2023){Berke}, {Murphy}, {Flynn}, \& {Liu}}]{Berke}
{Berke}, D.~A., {Murphy}, M.~T., {Flynn}, C., \& {Liu}, F. 2023, \mnras, 519, 1221, \dodoi{10.1093/mnras/stac2037}

\bibitem[{{Biazzo} {et~al.}(2022){Biazzo}, {D'Orazi}, {Desidera}, {Turrini}, {Benatti}, {Gratton}, {Magrini}, {Sozzetti}, {Baratella}, {Bonomo}, {Borsa}, {Claudi}, {Covino}, {Damasso}, {Di Mauro}, {Lanza}, {Maggio}, {Malavolta}, {Maldonado}, {Marzari}, {Micela}, {Poretti}, {Vitello}, {Affer}, {Bignamini}, {Carleo}, {Cosentino}, {Fiorenzano}, {Giacobbe}, {Harutyunyan}, {Leto}, {Mancini}, {Molinari}, {Molinaro}, {Nardiello}, {Nascimbeni}, {Pagano}, {Pedani}, {Piotto}, {Rainer}, \& {Scandariato}}]{Biazzo}
{Biazzo}, K., {D'Orazi}, V., {Desidera}, S., {et~al.} 2022, \aap, 664, A161, \dodoi{10.1051/0004-6361/202243467}

\bibitem[{{Bland-Hawthorn} \& {Gerhard}(2016)}]{solar_position_r}
{Bland-Hawthorn}, J., \& {Gerhard}, O. 2016, \araa, 54, 529, \dodoi{10.1146/annurev-astro-081915-023441}

\bibitem[{{Booth} \& {Owen}(2020)}]{Booth_engulf_gap}
{Booth}, R.~A., \& {Owen}, J.~E. 2020, \mnras, 493, 5079, \dodoi{10.1093/mnras/staa578}

\bibitem[{{Brown} {et~al.}(2011){Brown}, {Latham}, {Everett}, \& {Esquerdo}}]{KIC}
{Brown}, T.~M., {Latham}, D.~W., {Everett}, M.~E., \& {Esquerdo}, G.~A. 2011, \aj, 142, 112, \dodoi{10.1088/0004-6256/142/4/112}

\bibitem[{{Cassan} {et~al.}(2012){Cassan}, {Kubas}, {Beaulieu}, {Dominik}, {Horne}, {Greenhill}, {Wambsganss}, {Menzies}, {Williams}, {J{\o}rgensen}, {Udalski}, {Bennett}, {Albrow}, {Batista}, {Brillant}, {Caldwell}, {Cole}, {Coutures}, {Cook}, {Dieters}, {Dominis Prester}, {Donatowicz}, {Fouqu{\'e}}, {Hill}, {Kains}, {Kane}, {Marquette}, {Martin}, {Pollard}, {Sahu}, {Vinter}, {Warren}, {Watson}, {Zub}, {Sumi}, {Szyma{\'n}ski}, {Kubiak}, {Poleski}, {Soszynski}, {Ulaczyk}, {Pietrzy{\'n}ski}, \& {Wyrzykowski}}]{Cassan}
{Cassan}, A., {Kubas}, D., {Beaulieu}, J.~P., {et~al.} 2012, \nat, 481, 167, \dodoi{10.1038/nature10684}

\bibitem[{{Chambers}(2010)}]{Cham_TEABUND}
{Chambers}, J.~E. 2010, \apj, 724, 92, \dodoi{10.1088/0004-637X/724/1/92}

\bibitem[{{Chiba} \& {Beers}(2000)}]{Chiba_stackel}
{Chiba}, M., \& {Beers}, T.~C. 2000, \aj, 119, 2843, \dodoi{10.1086/301409}

\bibitem[{{Fischer} \& {Valenti}(2005)}]{Fischer}
{Fischer}, D.~A., \& {Valenti}, J. 2005, \apj, 622, 1102, \dodoi{10.1086/428383}

\bibitem[{{Gaia Collaboration} {et~al.}(2023){Gaia Collaboration}, {Montegriffo}, {Bellazzini}, {De Angeli}, {Andrae}, {Barstow}, {Bossini}, {Bragaglia}, {Burgess}, {Cacciari}, {Carrasco}, {Chornay}, {Delchambre}, {Evans}, {Fouesneau}, {Fr{\'e}mat}, {Garabato}, {Jordi}, {Manteiga}, {Massari}, {Palaversa}, {Pancino}, {Riello}, {Ruz Mieres}, {Sanna}, {Santove{\~n}a}, {Sordo}, {Vallenari}, {Walton}, {Brown}, {Prusti}, {de Bruijne}, {Arenou}, {Babusiaux}, {Biermann}, {Creevey}, {Ducourant}, {Eyer}, {Guerra}, {Hutton}, {Klioner}, {Lammers}, {Lindegren}, {Luri}, {Mignard}, {Panem}, {Pourbaix}, {Randich}, {Sartoretti}, {Soubiran}, {Tanga}, {Bailer-Jones}, {Bastian}, {Drimmel}, {Jansen}, {Katz}, {Lattanzi}, {van Leeuwen}, {Bakker}, {Casta{\~n}eda}, {Fabricius}, {Galluccio}, {Guerrier}, {Heiter}, {Masana}, {Messineo}, {Mowlavi}, {Nicolas}, {Nienartowicz}, {Pailler}, {Panuzzo}, {Riclet}, {Roux}, {Seabroke}, {Th{\'e}venin}, {Gracia-Abril}, {Portell}, {Teyssier}, {Altmann}, {Audard}, {Bellas-Velidis}, {Benson},
  {Berthier}, {Blomme}, {Busonero}, {Busso}, {C{\'a}novas}, {Carry}, {Cellino}, {Cheek}, {Clementini}, {Damerdji}, {Davidson}, {de Teodoro}, {Nu{\~n}ez Campos}, {Dell'Oro}, {Esquej}, {Fern{\'a}ndez-Hern{\'a}ndez}, {Fraile}, {Garc{\'\i}a-Lario}, {Gosset}, {Haigron}, {Halbwachs}, {Hambly}, {Harrison}, {Hern{\'a}ndez}, {Hestroffer}, {Hodgkin}, {Holl}, {Jan{\ss}en}, {Jevardat de Fombelle}, {Jordan}, {Krone-Martins}, {Lanzafame}, {L{\"o}ffler}, {Marchal}, {Marrese}, {Moitinho}, {Muinonen}, {Osborne}, {Pauwels}, {Recio-Blanco}, {Reyl{\'e}}, {Rimoldini}, {Roegiers}, {Rybizki}, {Sarro}, {Siopis}, {Smith}, {Sozzetti}, {Utrilla}, {van Leeuwen}, {Abbas}, {{\'A}brah{\'a}m}, {Abreu Aramburu}, {Aerts}, {Aguado}, {Ajaj}, {Aldea-Montero}, {Altavilla}, {{\'A}lvarez}, {Alves}, {Anderson}, {Anglada Varela}, {Antoja}, {Baines}, {Baker}, {Balaguer-N{\'u}{\~n}ez}, {Balbinot}, {Balog}, {Barache}, {Barbato}, {Barros}, {Bartolom{\'e}}, {Bassilana}, {Bauchet}, {Becciani}, {Berihuete}, {Bernet}, {Bertone}, {Bianchi}, {Binnenfeld},
  {Blanco-Cuaresma}, {Boch}, {Bombrun}, {Bouquillon}, {Bramante}, {Breedt}, {Bressan}, {Brouillet}, {Brugaletta}, {Bucciarelli}, {Burlacu}, {Butkevich}, {Buzzi}, {Caffau}, {Cancelliere}, {Cantat-Gaudin}, {Carballo}, {Carlucci}, {Carnerero}, {Casamiquela}, {Castellani}, {Castro-Ginard}, {Chaoul}, {Charlot}, {Chemin}, {Chiaramida}, {Chiavassa}, {Comoretto}, {Contursi}, {Cooper}, {Cornez}, {Cowell}, {Crifo}, {Cropper}, {Crosta}, {Crowley}, {Dafonte}, {Dapergolas}, {David}, {de Laverny}, {De Luise}, {De March}, {De Ridder}, {de Souza}, {de Torres}, {del Peloso}, {del Pozo}, {Delbo}, {Delgado}, {Delisle}, {Demouchy}, {Dharmawardena}, {Diakite}, {Diener}, {Distefano}, {Dolding}, {Enke}, {Fabre}, {Fabrizio}, {Faigler}, {Fedorets}, {Fernique}, {Figueras}, {Fournier}, {Fouron}, {Fragkoudi}, {Gai}, {Garcia-Gutierrez}, {Garcia-Reinaldos}, {Garc{\'\i}a-Torres}, {Garofalo}, {Gavel}, {Gavras}, {Gerlach}, {Geyer}, {Giacobbe}, {Gilmore}, {Girona}, {Giuffrida}, {Gomel}, {Gomez}, {Gonz{\'a}lez-N{\'u}{\~n}ez},
  {Gonz{\'a}lez-Santamar{\'\i}a}, {Gonz{\'a}lez-Vidal}, {Granvik}, {Guillout}, {Guiraud}, {Guti{\'e}rrez-S{\'a}nchez}, {Guy}, {Hatzidimitriou}, {Hauser}, {Haywood}, {Helmer}, {Helmi}, {Sarmiento}, {Hidalgo}, {H{\l}adczuk}, {Hobbs}, {Holland}, {Huckle}, {Jardine}, {Jasniewicz}, {Jean-Antoine Piccolo}, {Jim{\'e}nez-Arranz}, {Juaristi Campillo}, {Julbe}, {Karbevska}, {Kervella}, {Khanna}, {Kordopatis}, {Korn}, {K{\'o}sp{\'a}l}, {Kostrzewa-Rutkowska}, {Kruszy{\'n}ska}, {Kun}, {Laizeau}, {Lambert}, {Lanza}, {Lasne}, {Le Campion}, {Lebreton}, {Lebzelter}, {Leccia}, {Leclerc}, {Lecoeur-Taibi}, {Liao}, {Licata}, {Lindstr{\'o}m}, {Lister}, {Livanou}, {Lobel}, {Lorca}, {Loup}, {Madrero Pardo}, {Magdaleno Romeo}, {Managau}, {Mann}, {Marchant}, {Marconi}, {Marcos}, {Marcos Santos}, {Mar{\'\i}n Pina}, {Marinoni}, {Marocco}, {Marshall}, {Martin Polo}, {Mart{\'\i}n-Fleitas}, {Marton}, {Mary}, {Masip}, {Mastrobuono-Battisti}, {Mazeh}, {McMillan}, {Messina}, {Michalik}, {Millar}, {Mints}, {Molina}, {Molinaro}, {Moln{\'a}r},
  {Monari}, {Mongui{\'o}}, {Montero}, {Mor}, {Mora}, {Morbidelli}, {Morel}, {Morris}, {Muraveva}, {Murphy}, {Musella}, {Nagy}, {Noval}, {Oca{\~n}a}, {Ogden}, {Ordenovic}, {Osinde}, {Pagani}, {Pagano}, {Palicio}, {Pallas-Quintela}, {Panahi}, {Payne-Wardenaar}, {Pe{\~n}alosa Esteller}, {Penttil{\"a}}, {Pichon}, {Piersimoni}, {Pineau}, {Plachy}, {Plum}, {Poggio}, {Pr{\v{s}}a}, {Pulone}, {Racero}, {Ragaini}, {Rainer}, {Raiteri}, {Ramos}, {Ramos-Lerate}, {Re Fiorentin}, {Regibo}, {Richards}, {Rios Diaz}, {Ripepi}, {Riva}, {Rix}, {Rixon}, {Robichon}, {Robin}, {Robin}, {Roelens}, {Rogues}, {Rohrbasser}, {Romero-G{\'o}mez}, {Rowell}, {Royer}, {Rybicki}, {Sadowski}, {S{\'a}ez N{\'u}{\~n}ez}, {Sagrist{\`a} Sell{\'e}s}, {Sahlmann}, {Salguero}, {Samaras}, {Sanchez Gimenez}, {Sarasso}, {Schultheis}, {Sciacca}, {Segol}, {Segovia}, {S{\'e}gransan}, {Semeux}, {Shahaf}, {Siddiqui}, {Siebert}, {Siltala}, {Silvelo}, {Slezak}, {Slezak}, {Smart}, {Snaith}, {Solano}, {Solitro}, {Souami}, {Souchay}, {Spagna}, {Spina}, {Spoto},
  {Steele}, {Steidelm{\"u}ller}, {Stephenson}, {S{\"u}veges}, {Surdej}, {Szabados}, {Szegedi-Elek}, {Taris}, {Taylor}, {Teixeira}, {Tolomei}, {Tonello}, {Torra}, {Torra}, {Torralba Elipe}, {Trabucchi}, {Tsounis}, {Turon}, {Ulla}, {Unger}, {Vaillant}, {van Dillen}, {van Reeven}, {Vanel}, {Vecchiato}, {Viala}, {Vicente}, {Voutsinas}, {Wevers}, {Wyrzykowski}, {Yoldas}, {Yvard}, {Zhao}, {Zorec}, {Zucker}, \& {Zwitter}}]{GAIADR3}
{Gaia Collaboration}, {Montegriffo}, P., {Bellazzini}, M., {et~al.} 2023, \aap, 674, A33, \dodoi{10.1051/0004-6361/202243709}

\bibitem[{{Garc{\'\i}a P{\'e}rez} {et~al.}(2016){Garc{\'\i}a P{\'e}rez}, {Allende Prieto}, {Holtzman}, {Shetrone}, {M{\'e}sz{\'a}ros}, {Bizyaev}, {Carrera}, {Cunha}, {Garc{\'\i}a-Hern{\'a}ndez}, {Johnson}, {Majewski}, {Nidever}, {Schiavon}, {Shane}, {Smith}, {Sobeck}, {Troup}, {Zamora}, {Weinberg}, {Bovy}, {Eisenstein}, {Feuillet}, {Frinchaboy}, {Hayden}, {Hearty}, {Nguyen}, {O'Connell}, {Pinsonneault}, {Wilson}, \& {Zasowski}}]{ASPCAP}
{Garc{\'\i}a P{\'e}rez}, A.~E., {Allende Prieto}, C., {Holtzman}, J.~A., {et~al.} 2016, \aj, 151, 144, \dodoi{10.3847/0004-6256/151/6/144}

\bibitem[{Gardner {et~al.}(2006)Gardner, Mather, Clampin, Doyon, Greenhouse, Hammel, Hutchings, Jakobsen, Lilly, Long, Lunine, Mccaughrean, Mountain, Nella, Rieke, Rieke, Rix, Smith, Sonneborn, Stiavelli, Stockman, Windhorst, \& Wright}]{jwst}
Gardner, J.~P., Mather, J.~C., Clampin, M., {et~al.} 2006, Space Science Reviews, 123, 485, \dodoi{10.1007/s11214-006-8315-7}

\bibitem[{{Gibson} {et~al.}(2003){Gibson}, {Fenner}, {Renda}, {Kawata}, \& {Lee}}]{GCE}
{Gibson}, B.~K., {Fenner}, Y., {Renda}, A., {Kawata}, D., \& {Lee}, H.-c. 2003, \pasa, 20, 401, \dodoi{10.1071/AS03052}

\bibitem[{{Gonzalez}(1997)}]{Gonzal1997}
{Gonzalez}, G. 1997, \mnras, 285, 403, \dodoi{10.1093/mnras/285.2.403}

\bibitem[{{Gonzalez} {et~al.}(2010){Gonzalez}, {Carlson}, \& {Tobin}}]{Gonz_2010}
{Gonzalez}, G., {Carlson}, M.~K., \& {Tobin}, R.~W. 2010, \mnras, 407, 314, \dodoi{10.1111/j.1365-2966.2010.16900.x}

\bibitem[{{Gonz{\'a}lez Hern{\'a}ndez} {et~al.}(2013){Gonz{\'a}lez Hern{\'a}ndez}, {Delgado-Mena}, {Sousa}, {Israelian}, {Santos}, {Adibekyan}, \& {Udry}}]{Gon2013_nosignal}
{Gonz{\'a}lez Hern{\'a}ndez}, J.~I., {Delgado-Mena}, E., {Sousa}, S.~G., {et~al.} 2013, \aap, 552, A6, \dodoi{10.1051/0004-6361/201220165}

\bibitem[{{Gonz{\'a}lez Hern{\'a}ndez} {et~al.}(2011){Gonz{\'a}lez Hern{\'a}ndez}, {Israelian}, {Santos}, {Sousa}, {Delgado-Mena}, {Neves}, \& {Udry}}]{Gon2011}
{Gonz{\'a}lez Hern{\'a}ndez}, J.~I., {Israelian}, G., {Santos}, N.~C., {et~al.} 2011, in Astronomical Society of the Pacific Conference Series, Vol. 448, 16th Cambridge Workshop on Cool Stars, Stellar Systems, and the Sun, ed. C.~{Johns-Krull}, M.~K. {Browning}, \& A.~A. {West}, 879

\bibitem[{{Guerrero} {et~al.}(2021){Guerrero}, {Seager}, {Huang}, {Vanderburg}, {Garcia Soto}, {Mireles}, {Hesse}, {Fong}, {Glidden}, {Shporer}, {Latham}, {Collins}, {Quinn}, {Burt}, {Dragomir}, {Crossfield}, {Vanderspek}, {Fausnaugh}, {Burke}, {Ricker}, {Daylan}, {Essack}, {G{\"u}nther}, {Osborn}, {Pepper}, {Rowden}, {Sha}, {Villanueva}, {Yahalomi}, {Yu}, {Ballard}, {Batalha}, {Berardo}, {Chontos}, {Dittmann}, {Esquerdo}, {Mikal-Evans}, {Jayaraman}, {Krishnamurthy}, {Louie}, {Mehrle}, {Niraula}, {Rackham}, {Rodriguez}, {Rowden}, {Sousa-Silva}, {Watanabe}, {Wong}, {Zhan}, {Zivanovic}, {Christiansen}, {Ciardi}, {Swain}, {Lund}, {Mullally}, {Fleming}, {Rodriguez}, {Boyd}, {Quintana}, {Barclay}, {Col{\'o}n}, {Rinehart}, {Schlieder}, {Clampin}, {Jenkins}, {Twicken}, {Caldwell}, {Coughlin}, {Henze}, {Lissauer}, {Morris}, {Rose}, {Smith}, {Tenenbaum}, {Ting}, {Wohler}, {Bakos}, {Bean}, {Berta-Thompson}, {Bieryla}, {Bouma}, {Buchhave}, {Butler}, {Charbonneau}, {Doty}, {Ge}, {Holman}, {Howard}, {Kaltenegger}, {Kane},
  {Kjeldsen}, {Kreidberg}, {Lin}, {Minsky}, {Narita}, {Paegert}, {P{\'a}l}, {Palle}, {Sasselov}, {Spencer}, {Sozzetti}, {Stassun}, {Torres}, {Udry}, \& {Winn}}]{TOI}
{Guerrero}, N.~M., {Seager}, S., {Huang}, C.~X., {et~al.} 2021, \apjs, 254, 39, \dodoi{10.3847/1538-4365/abefe1}

\bibitem[{{Han} {et~al.}(2020){Han}, {Lee}, {Kim}, \& {Beers}}]{HanD}
{Han}, D.~R., {Lee}, Y.~S., {Kim}, Y.~K., \& {Beers}, T.~C. 2020, \apj, 896, 14, \dodoi{10.3847/1538-4357/ab919a}

\bibitem[{Harris {et~al.}(2020)Harris, Millman, van~der Walt, Gommers, Virtanen, Cournapeau, Wieser, Taylor, Berg, Smith, Kern, Picus, Hoyer, van Kerkwijk, Brett, Haldane, del R{\'{i}}o, Wiebe, Peterson, G{\'{e}}rard-Marchant, Sheppard, Reddy, Weckesser, Abbasi, Gohlke, \& Oliphant}]{numpy}
Harris, C.~R., Millman, K.~J., van~der Walt, S.~J., {et~al.} 2020, Nature, 585, 357, \dodoi{10.1038/s41586-020-2649-2}

\bibitem[{{Heiter} \& {Luck}(2003)}]{Heiter}
{Heiter}, U., \& {Luck}, R.~E. 2003, \aj, 126, 2015, \dodoi{10.1086/378366}

\bibitem[{{Hsu} {et~al.}(2019){Hsu}, {Ford}, {Ragozzine}, \& {Ashby}}]{Hsu}
{Hsu}, D.~C., {Ford}, E.~B., {Ragozzine}, D., \& {Ashby}, K. 2019, \aj, 158, 109, \dodoi{10.3847/1538-3881/ab31ab}

\bibitem[{{H{\"u}hn} \& {Bitsch}(2023)}]{Planet_formation_efficiency}
{H{\"u}hn}, L.~A., \& {Bitsch}, B. 2023, \aap, 676, A87, \dodoi{10.1051/0004-6361/202346604}

\bibitem[{Hunter(2007)}]{matplotlib}
Hunter, J.~D. 2007, Computing in Science \& Engineering, 9, 90, \dodoi{10.1109/MCSE.2007.55}

\bibitem[{{Johansen} {et~al.}(2009){Johansen}, {Youdin}, \& {Mac Low}}]{planet_formation_Johansen}
{Johansen}, A., {Youdin}, A., \& {Mac Low}, M.-M. 2009, \apjl, 704, L75, \dodoi{10.1088/0004-637X/704/2/L75}

\bibitem[{{J{\"o}nsson} {et~al.}(2020){J{\"o}nsson}, {Holtzman}, {Allende Prieto}, {Cunha}, {Garc{\'\i}a-Hern{\'a}ndez}, {Hasselquist}, {Masseron}, {Osorio}, {Shetrone}, {Smith}, {Stringfellow}, {Bizyaev}, {Edvardsson}, {Majewski}, {M{\'e}sz{\'a}ros}, {Souto}, {Zamora}, {Beaton}, {Bovy}, {Donor}, {Pinsonneault}, {Poovelil}, \& {Sobeck}}]{APG16}
{J{\"o}nsson}, H., {Holtzman}, J.~A., {Allende Prieto}, C., {et~al.} 2020, \aj, 160, 120, \dodoi{10.3847/1538-3881/aba592}

\bibitem[{{Kang} {et~al.}(2023){Kang}, {Lee}, {Kim}, \& {Beers}}]{Kang2023}
{Kang}, G., {Lee}, Y.~S., {Kim}, Y.~K., \& {Beers}, T.~C. 2023, \apjl, 954, L43, \dodoi{10.3847/2041-8213/ace32b}

\bibitem[{{Kawata} {et~al.}(2019){Kawata}, {Bovy}, {Matsunaga}, \& {Baba}}]{LSR}
{Kawata}, D., {Bovy}, J., {Matsunaga}, N., \& {Baba}, J. 2019, \mnras, 482, 40, \dodoi{10.1093/mnras/sty2623}

\bibitem[{{Kim} {et~al.}(2023){Kim}, {Lee}, {Beers}, \& {Kim}}]{Kim2023}
{Kim}, C., {Lee}, Y.~S., {Beers}, T.~C., \& {Kim}, Y.~K. 2023, Journal of Korean Astronomical Society, 56, 59, \dodoi{10.5303/JKAS.2023.56.1.59}

\bibitem[{{Kim} {et~al.}(2019){Kim}, {Lee}, \& {Beers}}]{Kim_stackel}
{Kim}, Y.~K., {Lee}, Y.~S., \& {Beers}, T.~C. 2019, \apj, 882, 176, \dodoi{10.3847/1538-4357/ab3660}

\bibitem[{{Kim} {et~al.}(2021){Kim}, {Lee}, {Beers}, \& {Koo}}]{Kim2021}
{Kim}, Y.~K., {Lee}, Y.~S., {Beers}, T.~C., \& {Koo}, J.-R. 2021, \apjl, 911, L21, \dodoi{10.3847/2041-8213/abf35e}

\bibitem[{{Lee} {et~al.}(2023){Lee}, {Lee}, {Kim}, {Beers}, \& {An}}]{Lee2023}
{Lee}, A., {Lee}, Y.~S., {Kim}, Y.~K., {Beers}, T.~C., \& {An}, D. 2023, \apj, 945, 56, \dodoi{10.3847/1538-4357/acb6f5}

\bibitem[{{Lee} {et~al.}(2019){Lee}, {Beers}, \& {Kim}}]{Lee2019}
{Lee}, Y.~S., {Beers}, T.~C., \& {Kim}, Y.~K. 2019, \apj, 885, 102, \dodoi{10.3847/1538-4357/ab4791}

\bibitem[{{Lindegren} {et~al.}(2021){Lindegren}, {Bastian}, {Biermann}, {Bombrun}, {de Torres}, {Gerlach}, {Geyer}, {Hern{\'a}ndez}, {Hilger}, {Hobbs}, {Klioner}, {Lammers}, {McMillan}, {Ramos-Lerate}, {Steidelm{\"u}ller}, {Stephenson}, \& {van Leeuwen}}]{Calibrationgdr3}
{Lindegren}, L., {Bastian}, U., {Biermann}, M., {et~al.} 2021, \aap, 649, A4, \dodoi{10.1051/0004-6361/202039653}

\bibitem[{{Liu} {et~al.}(2014){Liu}, {Asplund}, {Ramirez}, {Yong}, \& {Melendez}}]{WB_Liu_2014}
{Liu}, F., {Asplund}, M., {Ramirez}, I., {Yong}, D., \& {Melendez}, J. 2014, \mnras, 442, L51, \dodoi{10.1093/mnrasl/slu055}

\bibitem[{{Liu} {et~al.}(2021){Liu}, {Bitsch}, {Asplund}, {Liu}, {Murphy}, {Yong}, {Ting}, \& {Feltzing}}]{WB_Liu2021}
{Liu}, F., {Bitsch}, B., {Asplund}, M., {et~al.} 2021, \mnras, 508, 1227, \dodoi{10.1093/mnras/stab2471}

\bibitem[{{Liu} {et~al.}(2020){Liu}, {Yong}, {Asplund}, {Wang}, {Spina}, {Acu{\~n}a}, {Mel{\'e}ndez}, \& {Ram{\'\i}rez}}]{Liu2020}
{Liu}, F., {Yong}, D., {Asplund}, M., {et~al.} 2020, \mnras, 495, 3961, \dodoi{10.1093/mnras/staa1420}

\bibitem[{{Lodders}(2003)}]{lodd}
{Lodders}, K. 2003, \apj, 591, 1220, \dodoi{10.1086/375492}

\bibitem[{{Mackereth} {et~al.}(2019){Mackereth}, {Bovy}, {Leung}, {Schiavon}, {Trick}, {Chaplin}, {Cunha}, {Feuillet}, {Majewski}, {Martig}, {Miglio}, {Nidever}, {Pinsonneault}, {Aguirre}, {Sobeck}, {Tayar}, \& {Zasowski}}]{ageann}
{Mackereth}, J.~T., {Bovy}, J., {Leung}, H.~W., {et~al.} 2019, \mnras, 489, 176, \dodoi{10.1093/mnras/stz1521}

\bibitem[{{McDonough} \& {Sun}(1995)}]{terrestrial}
{McDonough}, W.~F., \& {Sun}, S.~s. 1995, Chemical Geology, 120, 223, \dodoi{10.1016/0009-2541(94)00140-4}

\bibitem[{McKinney {et~al.}(2010)}]{pandas}
McKinney, W., {et~al.} 2010, in Proceedings of the 9th Python in Science Conference, Vol. 445, Austin, TX, 51--56

\bibitem[{{Mel{\'e}ndez} {et~al.}(2009){Mel{\'e}ndez}, {Asplund}, {Gustafsson}, \& {Yong}}]{Mel}
{Mel{\'e}ndez}, J., {Asplund}, M., {Gustafsson}, B., \& {Yong}, D. 2009, \apjl, 704, L66, \dodoi{10.1088/0004-637X/704/1/L66}

\bibitem[{{Mel{\'e}ndez} {et~al.}(2012){Mel{\'e}ndez}, {Bergemann}, {Cohen}, {Endl}, {Karakas}, {Ram{\'\i}rez}, {Cochran}, {Yong}, {MacQueen}, {Kobayashi}, \& {Asplund}}]{Mel_2012_solar_simil}
{Mel{\'e}ndez}, J., {Bergemann}, M., {Cohen}, J.~G., {et~al.} 2012, \aap, 543, A29, \dodoi{10.1051/0004-6361/201117222}

\bibitem[{{Mishenina} {et~al.}(2021){Mishenina}, {Basak}, {Adibekyan}, {Soubiran}, \& {Kovtyukh}}]{Mishenina_2021}
{Mishenina}, T., {Basak}, N., {Adibekyan}, V., {Soubiran}, C., \& {Kovtyukh}, V. 2021, \mnras, 504, 4252, \dodoi{10.1093/mnras/stab1171}

\bibitem[{{Montalb{\'a}n} {et~al.}(2021){Montalb{\'a}n}, {Mackereth}, {Miglio}, {Vincenzo}, {Chiappini}, {Buldgen}, {Mosser}, {Noels}, {Scuflaire}, {Vrard}, {Willett}, {Davies}, {Hall}, {Nielsen}, {Khan}, {Rendle}, {van Rossem}, {Ferguson}, \& {Chaplin}}]{APOKA_age}
{Montalb{\'a}n}, J., {Mackereth}, J.~T., {Miglio}, A., {et~al.} 2021, Nature Astronomy, 5, 640, \dodoi{10.1038/s41550-021-01347-7}

\bibitem[{{Nibauer} {et~al.}(2021){Nibauer}, {Baxter}, {Jain}, {Van Saders}, {Beaton}, \& {Teske}}]{Niba}
{Nibauer}, J., {Baxter}, E.~J., {Jain}, B., {et~al.} 2021, \apj, 907, 116, \dodoi{10.3847/1538-4357/abd0f1}

\bibitem[{{Nissen}(2015)}]{Niss2015}
{Nissen}, P.~E. 2015, \aap, 579, A52, \dodoi{10.1051/0004-6361/201526269}

\bibitem[{{Nissen}(2016)}]{elemage2_nissen}
---. 2016, \aap, 593, A65, \dodoi{10.1051/0004-6361/201628888}

\bibitem[{{Nissen} {et~al.}(2020){Nissen}, {Christensen-Dalsgaard}, {Mosumgaard}, {Silva Aguirre}, {Spitoni}, \& {Verma}}]{elemage4_Nissen2020}
{Nissen}, P.~E., {Christensen-Dalsgaard}, J., {Mosumgaard}, J.~R., {et~al.} 2020, \aap, 640, A81, \dodoi{10.1051/0004-6361/202038300}

\bibitem[{{O'Connor} {et~al.}(2023){O'Connor}, {Bildsten}, {Cantiello}, \& {Lai}}]{engulf_common}
{O'Connor}, C.~E., {Bildsten}, L., {Cantiello}, M., \& {Lai}, D. 2023, \apj, 950, 128, \dodoi{10.3847/1538-4357/acd2d4}

\bibitem[{{Oh} {et~al.}(2018){Oh}, {Price-Whelan}, {Brewer}, {Hogg}, {Spergel}, \& {Myles}}]{WB_Oh}
{Oh}, S., {Price-Whelan}, A.~M., {Brewer}, J.~M., {et~al.} 2018, \apj, 854, 138, \dodoi{10.3847/1538-4357/aaab4d}

\bibitem[{Pedregosa {et~al.}(2011)Pedregosa, Varoquaux, Gramfort, Michel, Thirion, Grisel, Blondel, Prettenhofer, Weiss, Dubourg, Vanderplas, Passos, Cournapeau, Brucher, Perrot, \& Duchesnay}]{scikit-learn}
Pedregosa, F., Varoquaux, G., Gramfort, A., {et~al.} 2011, Journal of Machine Learning Research, 12, 2825

\bibitem[{{Pignatari} {et~al.}(2023){Pignatari}, {Trueman}, {Womack}, {Gibson}, {C{\^o}t{\'e}}, {Turrini}, {Sneden}, {Mojzsis}, {Stancliffe}, {Fong}, {Lawson}, {Keegans}, {Pilkington}, {Passy}, {Beers}, \& {Lugaro}}]{Pignatari_GCE}
{Pignatari}, M., {Trueman}, T. C.~L., {Womack}, K.~A., {et~al.} 2023, \mnras, 524, 6295, \dodoi{10.1093/mnras/stad2167}

\bibitem[{{Pinsonneault} {et~al.}(2001){Pinsonneault}, {DePoy}, \& {Coffee}}]{Pinso_engulf}
{Pinsonneault}, M.~H., {DePoy}, D.~L., \& {Coffee}, M. 2001, \apjl, 556, L59, \dodoi{10.1086/323531}

\bibitem[{{Pinsonneault} {et~al.}(2018){Pinsonneault}, {Elsworth}, {Tayar}, {Serenelli}, {Stello}, {Zinn}, {Mathur}, {Garc{\'\i}a}, {Johnson}, {Hekker}, {Huber}, {Kallinger}, {M{\'e}sz{\'a}ros}, {Mosser}, {Stassun}, {Girardi}, {Rodrigues}, {Silva Aguirre}, {An}, {Basu}, {Chaplin}, {Corsaro}, {Cunha}, {Garc{\'\i}a-Hern{\'a}ndez}, {Holtzman}, {J{\"o}nsson}, {Shetrone}, {Smith}, {Sobeck}, {Stringfellow}, {Zamora}, {Beers}, {Fern{\'a}ndez-Trincado}, {Frinchaboy}, {Hearty}, \& {Nitschelm}}]{APOKA2}
{Pinsonneault}, M.~H., {Elsworth}, Y.~P., {Tayar}, J., {et~al.} 2018, \apjs, 239, 32, \dodoi{10.3847/1538-4365/aaebfd}

\bibitem[{{Ram{\'\i}rez} {et~al.}(2009){Ram{\'\i}rez}, {Mel{\'e}ndez}, \& {Asplund}}]{Rami_2009}
{Ram{\'\i}rez}, I., {Mel{\'e}ndez}, J., \& {Asplund}, M. 2009, \aap, 508, L17, \dodoi{10.1051/0004-6361/200913038}

\bibitem[{{Ram{\'\i}rez} {et~al.}(2015){Ram{\'\i}rez}, {Khanal}, {Aleo}, {Sobotka}, {Liu}, {Casagrande}, {Mel{\'e}ndez}, {Yong}, {Lambert}, \& {Asplund}}]{WB_Rami}
{Ram{\'\i}rez}, I., {Khanal}, S., {Aleo}, P., {et~al.} 2015, \apj, 808, 13, \dodoi{10.1088/0004-637X/808/1/13}

\bibitem[{{Saffe} {et~al.}(2015){Saffe}, {Flores}, \& {Buccino}}]{WB_Saffe}
{Saffe}, C., {Flores}, M., \& {Buccino}, A. 2015, \aap, 582, A17, \dodoi{10.1051/0004-6361/201526644}

\bibitem[{{Saffe} {et~al.}(2017){Saffe}, {Jofr{\'e}}, {Martioli}, {Flores}, {Petrucci}, \& {Jaque Arancibia}}]{Saffe_engulf}
{Saffe}, C., {Jofr{\'e}}, E., {Martioli}, E., {et~al.} 2017, \aap, 604, L4, \dodoi{10.1051/0004-6361/201731430}

\bibitem[{{Santos, N. C.} {et~al.}(2004){Santos, N. C.}, {Israelian, G.}, \& {Mayor, M.}}]{santos}
{Santos, N. C.}, {Israelian, G.}, \& {Mayor, M.} 2004, A\&A, 415, 1153, \dodoi{10.1051/0004-6361:20034469}

\bibitem[{{Sch{\"o}nrich} {et~al.}(2010){Sch{\"o}nrich}, {Binney}, \& {Dehnen}}]{solar_motion}
{Sch{\"o}nrich}, R., {Binney}, J., \& {Dehnen}, W. 2010, \mnras, 403, 1829, \dodoi{10.1111/j.1365-2966.2010.16253.x}

\bibitem[{{Spina} {et~al.}(2016){Spina}, {Mel{\'e}ndez}, {Karakas}, {Ram{\'\i}rez}, {Monroe}, {Asplund}, \& {Yong}}]{Spina2016_elemage3}
{Spina}, L., {Mel{\'e}ndez}, J., {Karakas}, A.~I., {et~al.} 2016, \aap, 593, A125, \dodoi{10.1051/0004-6361/201628557}

\bibitem[{{Spina} {et~al.}(2018){Spina}, {Mel{\'e}ndez}, {Karakas}, {dos Santos}, {Bedell}, {Asplund}, {Ram{\'\i}rez}, {Yong}, {Alves-Brito}, {Bean}, \& {Dreizler}}]{Spina2018}
---. 2018, \mnras, 474, 2580, \dodoi{10.1093/mnras/stx2938}

\bibitem[{{Tautvai{\v{s}}ien{\.{e}}} {et~al.}(2022){Tautvai{\v{s}}ien{\.{e}}}, {Mikolaitis}, {Drazdauskas}, {Stonkut{\.{e}}}, {Minkevi{\v{c}}i{\={u}}t{\.{e}}}, {Pak{\v{s}}tien{\.{e}}}, {Kjeldsen}, {Brogaard}, {Chorniy}, {von Essen}, {Grundahl}, {Ambrosch}, {Bagdonas}, {Sharma}, \& {Viscasillas V{\'a}zquez}}]{Tau}
{Tautvai{\v{s}}ien{\.{e}}}, G., {Mikolaitis}, {\v{S}}., {Drazdauskas}, A., {et~al.} 2022, \apjs, 259, 45, \dodoi{10.3847/1538-4365/ac50b5}

\bibitem[{{Teske} {et~al.}(2016){Teske}, {Khanal}, \& {Ram{\'\i}rez}}]{WB_teske}
{Teske}, J.~K., {Khanal}, S., \& {Ram{\'\i}rez}, I. 2016, \apj, 819, 19, \dodoi{10.3847/0004-637X/819/1/19}

\bibitem[{{Tucci Maia} {et~al.}(2014){Tucci Maia}, {Mel{\'e}ndez}, \& {Ram{\'\i}rez}}]{WB_Tucci}
{Tucci Maia}, M., {Mel{\'e}ndez}, J., \& {Ram{\'\i}rez}, I. 2014, \apjl, 790, L25, \dodoi{10.1088/2041-8205/790/2/L25}

\bibitem[{{Udry} \& {Santos}(2007)}]{Udry_metal_host}
{Udry}, S., \& {Santos}, N.~C. 2007, \araa, 45, 397, \dodoi{10.1146/annurev.astro.45.051806.110529}

\bibitem[{Virtanen {et~al.}(2020)Virtanen, Gommers, Oliphant, Haberland, Reddy, Cournapeau, Burovski, Peterson, Weckesser, Bright, {van der Walt}, Brett, Wilson, Millman, Mayorov, Nelson, Jones, Kern, Larson, Carey, Polat, Feng, Moore, {VanderPlas}, Laxalde, Perktold, Cimrman, Henriksen, Quintero, Harris, Archibald, Ribeiro, Pedregosa, {van Mulbregt}, \& {SciPy 1.0 Contributors}}]{2020SciPy-NMeth}
Virtanen, P., Gommers, R., Oliphant, T.~E., {et~al.} 2020, Nature Methods, 17, 261, \dodoi{10.1038/s41592-019-0686-2}

\bibitem[{{Wang} {et~al.}(2022){Wang}, {Quanz}, {Yong}, {Liu}, {Seidler}, {Acu{\~n}a}, \& {Mojzsis}}]{Wang}
{Wang}, H.~S., {Quanz}, S.~P., {Yong}, D., {et~al.} 2022, \mnras, 513, 5829, \dodoi{10.1093/mnras/stac1119}

\bibitem[{{Xie} {et~al.}(2023){Xie}, {Zhu}, {Guo}, {Liu}, \& {L{\"u}}}]{engulf_Urich}
{Xie}, D., {Zhu}, C., {Guo}, S., {Liu}, H., \& {L{\"u}}, G. 2023, \mnras, 524, 3705, \dodoi{10.1093/mnras/stad2097}

\bibitem[{{Zink} {et~al.}(2020){Zink}, {Hardegree-Ullman}, {Christiansen}, {Petigura}, {Dressing}, {Schlieder}, {Ciardi}, \& {Crossfield}}]{Zink}
{Zink}, J.~K., {Hardegree-Ullman}, K.~K., {Christiansen}, J.~L., {et~al.} 2020, \aj, 160, 94, \dodoi{10.3847/1538-3881/aba123}

\end{thebibliography}
\bibliographystyle{aasjournal}

\end{document}